\documentclass[apjl]{emulateapj}
\usepackage{apjfonts}
\usepackage{graphicx}

\usepackage{amsmath}

\shorttitle{TOWARD UNDERSTANDING THE YOUNG SNRs INTERACTING WITH CLOUDS}
\shortauthors{T. INOUE ET. AL.}

\begin{document}

\title{
TOWARD UNDERSTANDING THE COSMIC-RAY ACCELERATION AT YOUNG SUPERNOVA REMNANTS INTERACTING WITH INTERSTELLAR CLOUDS: POSSIBLE APPLICATIONS TO RX J1713.7$-$3946
}
\author{Tsuyoshi Inoue\altaffilmark{1}, Ryo Yamazaki\altaffilmark{1}, Shu-ichiro Inutsuka\altaffilmark{2}, and Yasuo Fukui\altaffilmark{2}}
\altaffiltext{1}{Department of Physics and Mathmatics, Aoyama Gakuin University, Fuchinobe, Chuou-ku, Sagamihara 252-5258, Japan; inouety@phys.aoyama.ac.jp}
\altaffiltext{2}{Department of Physics, Graduate School of Science, Nagoya University, Furo-cho, Chikusa-ku, Nagoya 464-8602, Japan}

\begin{abstract}
Using three-dimensional magnetohydrodynamics simulations, we investigate general properties of a blast wave shock interacting with interstellar clouds.
The pre-shock cloudy medium is generated as a natural consequence of the thermal instability that simulates realistic clumpy interstellar clouds and their diffuse surrounding.
The shock wave that sweeps the cloudy medium generates a turbulent shell through the vorticity generations that are induced by shock-cloud interactions.
In the turbulent shell, the magnetic field is amplified as a result of turbulent dynamo action.
The energy density of the amplified magnetic field can locally grow comparable to the thermal energy density, particularly at the transition layers between clouds and the diffuse surrounding.
In the case of a young supernova remnant (SNR) with a shock velocity $\gtrsim 10^{3}$ km s$^{-1}$, the corresponding strength of the magnetic field is approximately 1 mG.
The propagation speed of the shock wave is significantly stalled in the clouds because of the high density, while the shock maintains a high velocity in the diffuse surrounding.
In addition, when the shock wave hits the clouds, reflection shock waves are generated that propagate back into the shocked shell.
From these simulation results, many observational characteristics of a young SNR RX J1713.7$-$3946 that is suggested to be interacting with molecular clouds, can be explained as follows:
The reflection shocks can accelerate particles in the turbulent downstream region where the magnetic field strength reaches 1mG, which causes  short-time variability of synchrotron X-rays.
Since the shock velocity is stalled locally in the clouds, the temperature in the shocked cloud is suppressed far below 1 keV.
Thus, thermal X-ray line emission would be faint even if the SNR is interacting with molecular clouds.
We also find that the photon index of the $\pi^0$-decay gamma rays generated by cosmic-ray protons can be 1.5 (corresponding energy flux is $\nu F_{\nu}\propto \nu^{0.5}$), because the penetration depth of high-energy particles into the clumpy clouds depends on their energy.
This suggests that, if we rely only on the spectral study, the hadronic gamma-ray emission is indistinguishable from the leptonic inverse Compton emission.
We propose that the spatial correlation of the gamma-ray, X-ray, and CO line emission regions can be conclusively used to understand the origin of gamma rays from RX J1713.7$-$3946.
\end{abstract}


\section{Introduction}
Supernova remnants (SNRs) are believed to be the sites of Galactic cosmic-ray acceleration through a diffusive shock acceleration mechanism (DSA; Bell 1978, Blandford \& Ostriker 1978, Blandford \& Eichler 1987), and multi-wavelength nonthermal emissions from SNRs caused by accelerated particles have been detected (e.g., Koyama et al. 1995, Aharonian et al. 2008).
However the detailed process of DSA and the emission mechanism of SNRs are still a matter of debate.
Recent observations have emphasized the importance of the interaction between SNRs and interstellar clouds.
The {\it Fermi Gamma-ray Space Telescope} revealed gamma-ray emissions from middle-aged SNRs interacting with molecular clouds (Abdo et al. 2009, 2010a, 2010b, 2010c).
In addition, in a young SNR, RX J1713.7$-$3946 (or G347.3$-$0.5), spatial correlation between molecular clouds and X-/gamma-ray emissions has been reported (Fukui et al. 2003, 2008, 2011, Moriguchi et al. 2005, Sano et al. 2010).

When we consider the interaction between a SNR and interstellar clouds, we should take into account the highly inhomogeneous structure of clouds (see also Laming 2001a, b for the effects of circumstellar density inhomogeneity).
Recent numerical simulations have shown that interstellar clouds are formed as a complex of clumps fragmented by the thermal instability that are embedded in diffuse gas (Koyama \& Inutsuka 2002; Hennebelle et al. 2008; Inoue \& Inutsuka 2008; 2009; Banerjee et al. 2009; Heitsch et al. 2009; V\'azquez-Semadeni et al. 2006, Audit \& Hennebelle 2010).
Furthermore, in molecular clouds, supersonic turbulence is always observed as a supra-thermal line width of molecular line emissions (e.g., Larson 1981, Heyer \& Brunt 2004) that inevitably generate highly inhomogeneous structures by shock compressions (see, e.g., MacLow \& Klessen 2004).
By using two-dimensional magnetohydrodynamics (MHD) simulations, Inoue, Yamazaki \& Inutsuka (2009) demonstrated that the interaction between a strong shock wave and a cloudy inhomogeneous medium generates a turbulent SNR shell.
The turbulence induced by the shock-cloud interactions amplifies the magnetic field through turbulent dynamo actions (see also, Balsara et al. 2001; Giacalone \& Jokipii 2007), which potentially account for the year-scale short-time variability of X-rays in RX J1713.7$-$3946 (Uchiyama et al. 2007) as well as the spatial scale of the regions with the short-term variation.
In the follow-up study by Inoue, Yamazaki \& Inutsuka (2010), a similar but more realistic three-dimensional simulation was performed in which it was found that the shock-cloud interactions cause reflected shock waves in the SNR that were shown to reproduce the broken power-law cosmic-ray spectrum observed in the above-mentioned middle-aged SNRs (see also Malkov et al. 2010, Ohira et al. 2010, Uchiyama et al. 2010).
The shock velocity examined in Inoue, Yamazaki \& Inutsuka (2010) is much smaller than that expected in young SNRs because we discussed middle-aged SNRs and focused only on the physical properties of the reflected shock waves.
In this paper, we study the shock-cloud interaction in more detail and examine its influences especially on young SNRs aged $\sim 1000$ years such as RX J1713.7$-$3946.

A number of observational and theoretical efforts have been devoted to understand the emission mechanisms of RX J1713.7$-$3946 (Enomoto et al. 2002, Butt et al. 2002, Aharonian et al. 2006, 2007, Uchiyama et al. 2007, Tanaka et al. 2008, Kats \& Waxman 2008, Plaga 2008, Berezhko \& V$\ddot{\mbox{o}}$lk 2008, Acero et al. 2009, Fang et al. 2009, Morlino et al. 2009, Yamazaki et al. 2009, Zirakashvili \& Aharonian 2010, Ellison et al. 2010, Zhang \& Yang 2011, Abdo et al. 2011).
Based on the recent gamma-ray observation using the Fermi space telescope, it is suggested that the spectral energy distribution prefers the emission mechanism of the inverse Compton scattering of the cosmic microwave background (CMB) photons by high-energy electrons accelerated by the shock wave (Abdo et al. 2011).
Theoretical modeling of RX J1713.7$-$3946 also supports the leptonic origin of the gamma-ray emission;
Ellison et al. (2010)  used one-dimensional hydrodynamics simulations of a supernova blast wave to demonstrate that the hadronic model of gamma-ray emission fails to reproduce observed X-ray emission due to an overproduction of thermal X-ray line emission.
However, the leptonic model of gamma-ray emission from RX J1713.7$-$3946 also has problems.
If the gamma-ray emission is leptonic, the magnetic field strength should be $B\sim 10$ $\mu$G from the ratio of synchrotron X-ray and inverse Compton gamma-ray fluxes, while the short-time variability observed in some X-ray bright regions indicates that the magnetic field strength can be as large as $B\sim 1$ mG, and the thickness of the X-ray filaments indicates $B\sim100$ $\mu$G around the shock front (Ballet 2006, Uchiyama et al. 2007, Tanaka et al. 2008, Acero et al. 2009), which is similar to other young SNRs (Vink \& Laming 2003, Bamba et al. 2003, 2005).

The lack of thermal X-ray line emission from RX J1713.7$-$3946 is apparently very crucial when we develop a shock-cloud interaction model of young SNRs, because the shocked dense clouds are thought to emit numerous thermal X-ray lines.
In this paper, however, using three-dimensional MHD simulations, we show that the clouds shocked by a supernova blast wave do not emit thermal X-ray lines, because the transmitted shock wave stalls in the dense clouds.
This may resolve the problem that the thermal X-ray line emission is substantially suppressed in RX J1713.7$-$3946 despite the suggested interaction with dense molecular clouds.
We also show that the photon index of the hadronic gamma rays can be $p-1/2=1.5$ for $p=2$, where $p$ is the spectrum index of accelerated protons, which is consistent with the recent gamma-ray observation by Abdo et al. (2011), if the interacting molecular clouds are clumpy as demanded from theoretical arguments, numerical simulations, and observational facts mentioned above.
This indicates that it is difficult to distinguish the leptonic and hadronic gamma-ray emissions from the spectral study alone.
We propose the spatial correlation of the gamma-ray, X-ray, and CO line emission regions as a conclusive tool to understand the origin of gamma rays from RX J1713.7$-$3946.

The organization of the paper is as follows:
In \S 2, we briefly explain why we consider inhomogeneous clouds and their surrounding and provide numerical settings of our simulations.
The results of the simulations are shown in \S 3.
In \S 4, we discuss implications of the simulations and their application to the young SNR RX J1713.7$-$3946.
Finally in \S 5, we summarize our findings.

\section{Setup of Simulations}
\subsection{Thermal Instability in the ISM}

\begin{figure}[t]
\epsscale{1.}
\plotone{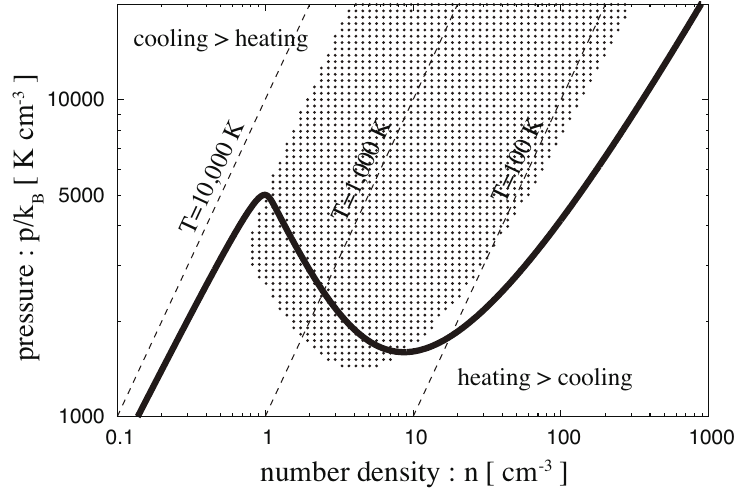}
\caption{
Thermal equilibrium curve in typical ISM in which cooling and heating are in balance ({\it thick solid}).
A fitting cooling/heating function given in Koyama \& Inutsuka (2002) is used to calculate the curve.
The dotted region is the thermally unstable region in which the Balbus instability criterion eq. (\ref{bulc}) is satisfied.
Dashed lines are isotherms of $T=10^2$, $10^3$, and $10^4$ K.
}
\label{f1}
\end{figure}

In this paper, we consider inhomogeneous interstellar clouds as a pre-shock medium.
In the following, we briefly explain how inhomogeneity is imprinted in interstellar clouds.
It is widely known that the interstellar medium (ISM) is an energy-open system due to the radiative cooling and heating that make the ISM a thermally bistable medium (Field et al. 1969; Wolfire et al. 1995; 2003; Koyama \& Inutsuka 2000).
Fig. \ref{f1} shows the thermal equilibrium curve in a typical ISM in which cooling and heating are in balance (Koyama \& Inutsuka 2002).
Two roughly isothermal states with $T\sim 10^2$ K and $10^4$ K correspond, respectively, to the phases of interstellar cloud and diffuse intercloud gas that are connected by a thermally unstable equilibrium.
Below the equilibrium curve, cooling dominates heating, and above it, heating dominates cooling.
The thermal instability is the most promising formation mechanism of interstellar clouds that is driven by runaway cooling.
The instability criterion is given by the condition (Balbus 1995):
\begin{equation}\label{bulc}
\left\{ \frac{\partial }{\partial T} \left(\frac{{\cal L}}{T}\right) \right\}_{p}<0,
\end{equation}
where ${\cal L}(n,T)$ is the net cooling rate per unit mass.
The dotted region in Fig. \ref{f1} is the thermally unstable region in which density inhomogeneities grow exponentially toward stable equilibria (see, Field 1965, Schwarz et al. 1972, and Koyama \& Inutsuka 2000 for linear stability analyses under various dynamical conditions).
From Fig. \ref{f1}, it is clear that, during any formation process of clouds from the diffuse intercloud gas, the gas always experiences thermal instability.
In a typical ISM, the timescale of the thermal instability is given by the cooling timescale that is $t_{\rm cool}\sim 1$ Myr, and the most unstable scale of the thermal instability is $l_{\rm TI}\sim1$ pc.
The nonlinear growth of the thermal instability toward the cold phase (condensation) generates a cold clump whose scale is much smaller than $l_{\rm TI}$.
This is the reason why recent numerical simulations of molecular and HI clouds formation show the generation of clouds as a complex of cloudlets (Koyama \& Inutsuka 2002; Hennebelle et al. 2008; Inoue \& Inutsuka 2008; 2009; Banerjee et al. 2009; Heitsch et al. 2009; Audit \& Hennebelle 2010).
It is also known from the comparisons of the growth timescales of various hydrodynamic instabilities that the thermal instability dominates the dynamics of cloud formation (Heitsch et al. 2008).
The simulations of the global ISM under the influence of supernovae have also pointed out that the multiphase structure in the ISM is regulated by the dynamical process of the thermal instability (MacLow et al. 2005, de Avillez \& Breitschwerdt 2005).
Observationally, Sakamoto \& Sunada (2003) found, at the envelope in the Taurus molecular cloud, that the cloud is indeed composed of the small-scale clumps where the thermal instability plays a role.

In molecular clouds, there is an additional generator of density inhomogeneity other than the thermal instability.
Molecular line emissions from clouds are always observed with supra-thermal line widths that are considered to be the outcome of supersonic turbulence (Larson 1981, Heyer \& Brunt 2004, Mac Low \& Klessen 2004).
The supersonic turbulence inevitably introduces density fluctuations due to the shock compressions of converging flows and rarefactions by diverging flows, even if we artificially set up a uniform molecular cloud initially (see, e.g., Padoan \& Nordlund 1999).

\subsection{Generation of Cloudy ISM through Thermal Instability}
From the above-mentioned understanding of interstellar clouds, we employ a cloudy ISM formed by the thermal instability as an ambient medium of a SNR.
We solve the three-dimensional ideal MHD equations with interstellar cooling, heating, and thermal conduction:
\begin{eqnarray}
&&\frac{\partial\rho}{\partial t}+\vec{\nabla}\cdot(\rho\,\vec{v})=0,\nonumber\\
&&\frac{\partial\rho\,\vec{v}}{\partial t}+\vec{\nabla}\cdot(p+\frac{B^2}{8\,\pi}+\rho\,\vec{v}\otimes\vec{v}-\frac{\vec{B}\otimes\vec{B}}{4\,\pi})=0,\nonumber\\
&& \frac{\partial e}{\partial t}+\vec{\nabla}\cdot\{(e+p+\frac{B^2}{8\,\pi})\,\vec{v}-\frac{\vec{B}\cdot\vec{v}}{4\,\pi}\vec{B}\}=\vec{\nabla}\cdot\kappa\vec{\nabla} T-\rho\,{\cal L}(n,T),\nonumber\\
&&\frac{\partial \vec{B}}{\partial t}=\vec{\nabla}\times(\vec{v}\times\vec{B}),\nonumber\\
&&e=\frac{p}{\gamma-1}+\frac{\rho\,v^2}{2}+\frac{B^2}{8\,\pi},\nonumber
\end{eqnarray}
where $\kappa$ is the thermal conductivity and ${\cal L}(n,T)$ is the net cooling rate per unit mass.
We employ a net cooling function that is obtained by fitting various line-emission coolings (Ly-$\alpha$, CII $158 \mu$m, OI $63 \mu$m, etc.) and photoelectric heating by dust (Koyama \& Inutsuka 2002), which can adequately describe the effects of cooling and heating roughly in the temperature range $10$ K $\lesssim T \lesssim 10^4$ K.
We impose the ideal gas equation of state and the adiabatic index $\gamma=5/3$.
Since the thermal conduction determines the most unstable scale of the thermal instability (Field 1965), it should be taken into account during cloud formation.
Because the medium is in a weakly ionized state, the isotropic thermal conductivity due to the neutral atomic collisions $\kappa = 2.5\times 10^3$ erg cm$^{-1}$ s$^{-1}$ K$^{-1}$ is used (Parker 1953).
The numerical technique used to solve the basic equations is a combination of a second-order Godunov-type finite volume scheme (Sano et. al. 1999) and a second-order consistent method of characteristics with constrained transport algorithm (Clarke 1996).
The former technique enables us to sharply capture a shock wave (van Leer 1979), and the later allows us to integrate the induction equation without breaking the divergence free condition of magnetic field (Evans \& Hawley 1988) and to stably follow the strong magnetic field amplification due to turbulence (Clarke 1996).

\begin{figure}[t]
\epsscale{1.}
\plotone{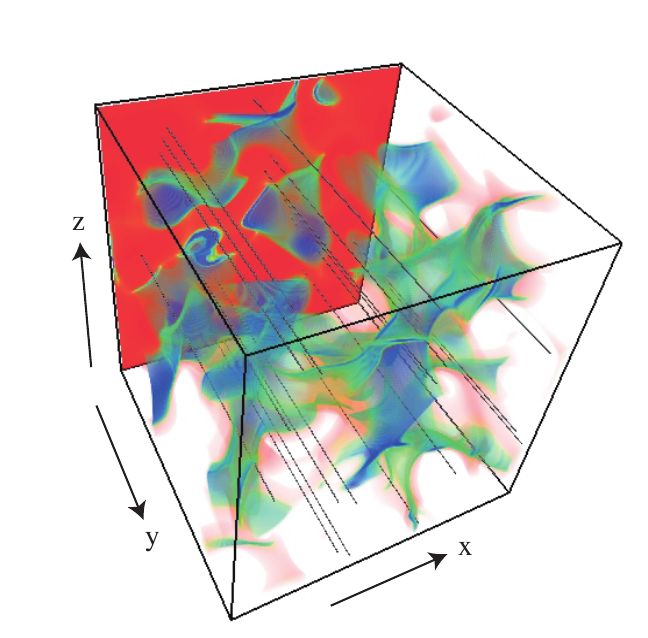}
\caption{
Number density volume rendering of the resulting cloudy medium as a consequence of the thermal instability after 3.0 Myr of evolution (a few cooling times).
The number density map in the $y=0.0$ pc plane is overplotted.
Regions in green and blue indicate the density $n\sim 10$ cm$^{-3}$ and $n\gtrsim 30$ cm$^{-3}$, respectively, and the region in red shows the diffuse intercloud gas with $n\lesssim 1$ cm$^{-3}$.
Magnetic field lines are represented as gray lines.
}
\label{f2}
\end{figure}

We use a cubic numerical domain whose side lengths are 2 pc with the resolution of $\Delta x =2$ pc$/1024=1.95\times10^{-3}$ pc in which periodic boundary conditions are imposed.
 As the initial conditions, we choose a thermally unstable equilibrium state with random density fluctuations whose mean density and thermal pressure are $n=2.0$ cm$^{-3}$ and $p/k_{\rm B}=2887$ K cm$^{-3}$, respectively.
The uniform magnetic field oriented +$y$ direction is imposed whose strength is 5.0 $\mu$G, which is believed to be the average strength in the ISM (Beck 2000).
Inoue \& Inutsuka (2008, 2009) showed that such a thermally unstable medium can be ubiquitously expected as an initial condition of interstellar clouds.

Fig. \ref{f2} shows the number density volume rendering of the resulting cloudy medium as a consequence of the thermal instability after 3.0 Myr of evolution (a few cooling times).
The number density map in the $y=0.0$ pc plane is overplotted.
Regions in green and blue indicate the density $n\sim 10$ cm$^{-3}$ and $n\gtrsim 30$ cm$^{-3}$, respectively, and the region in red shows the diffuse intercloud gas with $n\lesssim 1$ cm$^{-3}$.
Magnetic field lines are represented as gray lines.
The condensations driven by the thermal instability to form clouds arise along the magnetic field lines, since the motion perpendicular to the field is easily stopped due to the magnetic pressure.
This results in a formation of sheet-like clouds whose thicknesses are essentially determined by the most unstable scale of the thermal instability ($\sim$ 1 pc) times the compression ratio of the condensation ($\sim 0.1$) that gives $\sim 0.1$ pc, and whose length is roughly given by the most unstable scale $\sim 1$ pc.
Note that the intercloud gas is also thermally unstable due to runaway heating that evolves toward the diffuse gas phase.
The typical densities and temperatures of the clouds are $n_c\simeq 40$ cm$^{-3}$ and $T_c\simeq 100$ K, and those of the diffuse intercloud gas are $n_d\simeq 1$ cm$^{-3}$ and $T_d\simeq 5,000$ K.
A more detailed description of the evolution of the thermal instability can be found, e.g., in Inoue et al. (2007) and Inoue \& Inutsuka (2008, 2009).

In this simulation, the formed clouds correspond to HI clouds and their volume filling factor (where $n>10$ cm$^{-3}$) is 1.9\%.
In \S 3 and \S 4, we discuss how cloud density and cloud filling factor affect the dynamics of SNR.
We stress that even in the formation of denser molecular cloud, dense clumps are generated by the thermal instability and the clumps are embedded in the diffuse gas (see, e.g., Hennebelle et al. 2008; Banerjee et al. 2009).

\subsection{Induction of Shock Wave}
To study the formation of SNRs from the cloudy ISM, we induce a strong shock wave by setting a high-pressure hot gas with $p_{\rm h}/k_{\rm B}=1.0\times10^9$ and $n_{\rm h}=0.1$ cm$^{-3}$ at one of the boundaries of the simulation domain.
We examine two cases of shock induction from the $x=0$ surface and the $y=0$ surface, which correspond to the perpendicular shock and the parallel shock, respectively.
The resulting average propagation speeds of the induced shock waves are 2429 km s$^{-1}$ and 2330 km s$^{-1}$ for the perpendicular and parallel shock cases, respectively, which correspond to the supernova blast wave shocks with an age of $\sim 10^3$ yr.
For the perpendicular (parallel) shock case, we set the periodic boundary condition at the $y$ ($x$) and $z$ boundaries and the free boundary condition at the $x=2$ ($y=2$) pc boundary.
This numerical setting is very similar to the simulation performed in Inoue et al. (2010) in which only the perpendicular shock with a much smaller shock velocity ($v_{\rm sh}\sim 500$ km s$^{-1}$) is presented.

In contrast to the stage of setting up the initial condition, we omit the effect of cooling in the simulation of shock propagation, since the cooling timescale in shocked diffuse gas is larger than the timescale of the shock crossing time.
As for the cooling in the shocked cloud, it can also be also neglected in our choice of initial medium.
According to the fitting line emission cooling rate given by Gaetz et al. (1987), the cooling timescale for shocked cloud is estimated to be
\begin{eqnarray}
t_{\rm cool}&\simeq& 3\times10^3\,\left( \frac{v_{sh}}{3000\,\mbox{km s}^{-1}}\right)^{3.5}\,\left( \frac{n_{c}}{100\,\mbox{cm}^{-3}}\right)^{-2.75}\,\nonumber \\
&&\times \left( \frac{n_{d}}{1\,\mbox{cm}^{-3}}\right)^{1.75}\,\mbox{yr},
\end{eqnarray}
where $v_{sh}$ is the shock speed in the diffuse medium and we have used the fact that the shock velocity in clouds is stalled by a factor of $\sqrt{n_d/n_c}$ (see, \S 3.1 below).
This cooling timescale is longer than the age of typical young SNR $\lesssim$ 1,000 yr.
However, as we discuss in \S 4.3 and \S 4.4, the cloud density that seems to be interacting with the shock in RXJ1713.7$-$3946 may be much larger (typically $n_c \gtrsim 1,000$ cm$^{-3}$).
In that case, the timescale of cooling becomes much smaller than the age of the SNR.
As we have shown in Appendix A, the dynamics outside the shocked cloud marginally depends on the effect of cooling in the shocked cloud.
Thus, even if the effect of cooling in shocked cloud is very effective (i.e., $\gamma=1$ in the cloud), we can expect the similar results outside the very dense regions where generation of turbulence and magnetic field amplification take place (see \S 3.1 and \S 3.2).
We also omit the effect of thermal conduction, since the shocked gas becomes a fully ionized gas in which the gyro radius for thermal protons is much smaller than the resolution of our simulation:
\begin{eqnarray}
l_{\rm g, p}&\simeq&3 \times 10^{9}\,\left( \frac{p/k_{\rm B}}{10^9\,\mbox{K cm}^{-3}} \right)^{1/2}\,\left( \frac{n}{10\,\mbox{cm}^{-3}} \right)^{-1/2} \nonumber \\
&&\times\,\left( \frac{B}{5\,\mu\mbox{G}} \right)^{-1}\,\mbox{cm}\ll \Delta x\simeq 6\times 10^{15}\,\mbox{cm}.
\end{eqnarray}

Note that, since the basic MHD equations are integrated in conservative fashion, our code can accurately handle high Mach number shocks.
Since we perform this shock propagation simulations in the upstream-rest-flame, the kinetic and magnetic energy does not dominate the thermal energy everywhere in the computational domain that enable us to perform the simulations of a very high Mach number shock propagation even in the use of the total energy conservative code.

\section{Results}
\subsection{Generation of Turbulent SNR}\label{Sturb}
\begin{figure*}[t]
\epsscale{1.}
\plotone{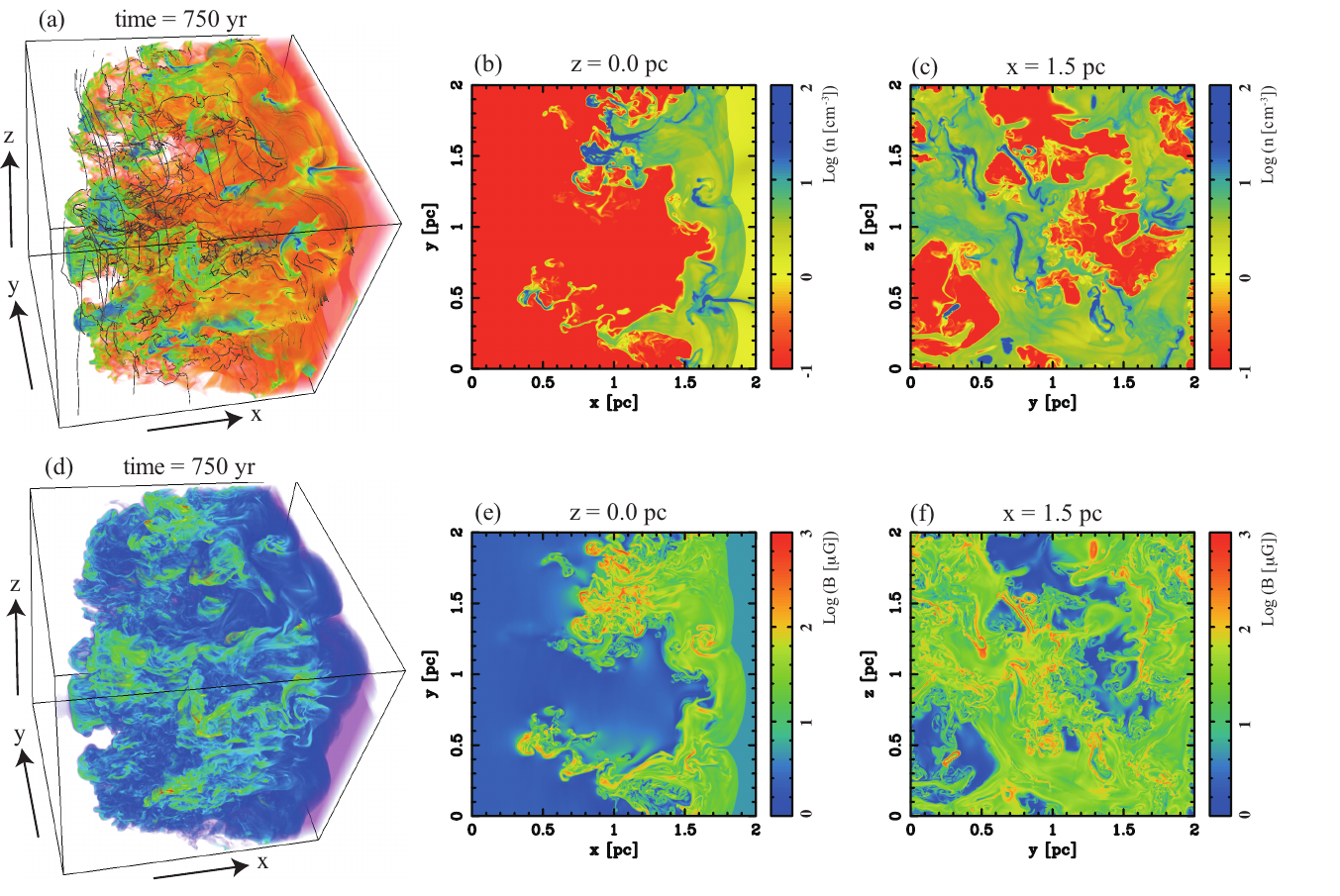}
\caption{
Result of the perpendicular shock case at $t = 750$ yr after the shock injection.
Panel (a): number density volume rendering.
Regions in green and blue indicate the density $n\sim 10$ cm$^{-3}$ and $n\gtrsim 30$ cm$^{-3}$, respectively, and the regions in warm colors show the shocked diffuse gas with $n\lesssim 4$ cm$^{-3}$.
Magnetic field lines are represented as gray lines.
Panel (b): two-dimensional number density slice at $z=0.0$ pc.
Panel (c): slice of number density at $x=1.5$ pc.
Panel (d): volume rendering of magnetic field strength.
Regions in blue, green and red indicate the regions with $B\lesssim 100$ $\mu$G, $B\gtrsim 100$ $\mu$G, and $B\gtrsim 500$ $\mu$G, respectively.
Panel (e): slice of magnetic field strength at $z = 0.0$ pc.
Panel (f): slice of magnetic field strength at $x = 1.5$ pc.
}
\label{f3}
\end{figure*}

\begin{figure*}[t]
\epsscale{1.}
\plotone{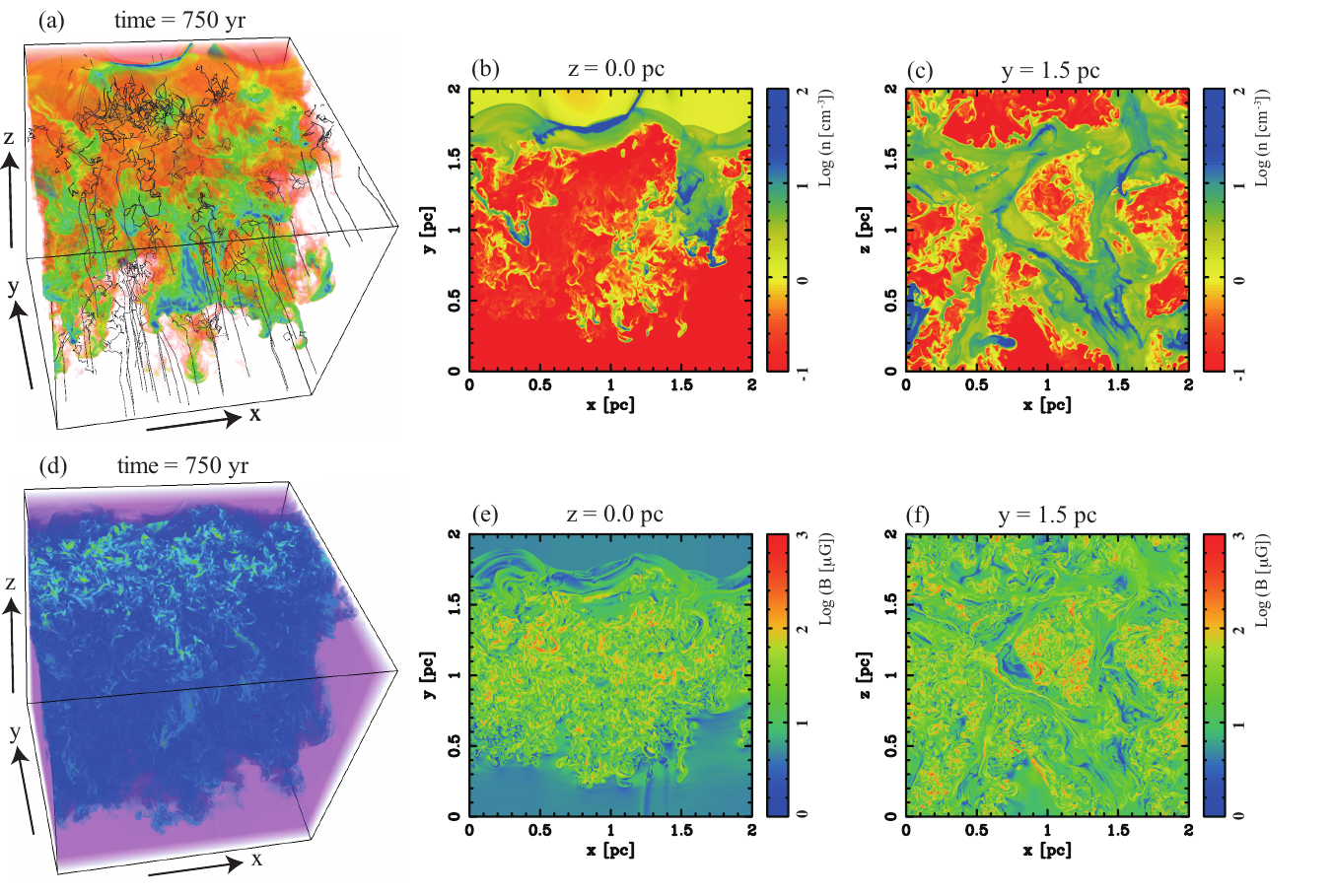}
\caption{
Same as Fig. \ref{f3} but for the parallel shock case.
}
\label{f4}
\end{figure*}

Propagation of the shock wave compresses and piles up the cloudy medium and forms a turbulent shocked slab.
Figs. \ref{f3} and \ref{f4} show density and magnetic field strength structures of the perpendicular and the parallel shock cases at $t=750$ yr after the shock injection, respectively.
When the shock wave hits a cloud, a transmitted shock wave propagates into the cloud whose propagation speed is slower than the shock velocity in the diffuse intercloud gas by a factor of $\sqrt{n_{\rm i}/n_{\rm c}}$ (see Appendix A for details), where $n_{\rm i}$ and $n_{\rm c}$ are the number densities of intercloud gas and cloud, respectively.
The stalling of the shock wave can be easily understood as follows: In the case of a blast wave, the pressure behind the shock becomes roughly isobaric, except with fluctuations of an order of magnitude (see \S3.4 below), and the shock velocity in the pre-shock rest frame is essentially determined by the post-shock sound speed.
Thus, the shock propagation speed becomes inversely proportional to the square root of pre-shock density.
As a result of the stall, the shock front is deformed when it interacts with dense clumps.
In the panels (b) of Figs. \ref{f3} and \ref{f4}, the deformations of the shock waves are clearly observed.
The deformation of the shock front due to a pre-shock density fluctuation is known as the Richtmyer-Meshkov instability (see, Brouillette 2002, Nishihara et al. 2010 for reviews).
The deformed shock wave leaves vorticity in the post-shock flow, even if the pre-shock is static (Crocco's theorem; see also Kida \& Orszag 1990, Zabusky 1999) that drives turbulence.

In addition to this curved shock effect, the baroclinic effect due to the misalignment of pressure and density gradients also generates vorticity.
In the simulations, the baroclinic effect is significant at the transition layer between the cloud and diffuse gas where density gradient is very large.
The vorticity generated by the baroclinic effect is the strong velocity shear flows between the shocked cloud and the diffuse gas.
In other words, the velocity difference in the shocked cloud and shocked diffuse gas generates the velocity shear, since the shock speed is stalled in the cloud while the shock and post-shock gas around the cloud continue to propagate with a faster velocity.

The velocity dispersion of the shocked gas for the perpendicular shock case is 705 km s$^{-1}$ and that for the parallel shock case is 738 km s$^{-1}$.
The corresponding Mach numbers are, respectively, $\langle {\cal M}^2 \rangle^{1/2}$=0.61 and 0.62.
Note that these dispersions mainly represent the velocity dispersions of diffuse gas.
Here we have defined the shocked gas as the gas with $n>0.5$ cm$^{-3}$ and $p/k_{\rm B}> 10^4$ K cm$^{-3}$.
The former condition excludes the thin hot gas injected from the boundary and the later condition excludes the pre-shock gas.
In what follows, we use the above two conditions to select the shocked gas.

As we mentioned in the previous section, most of the volume in the initial medium is filled by diffuse gas.
Thus, the global shock speed is essentially determined by the shock propagation speed in the diffuse gas.
This indicates that even if a SNR is interacting with interstellar clouds, the global expansion rate of SNR is determined by the shock propagation speed in the diffuse gas.

\subsection{Magnetic Field Amplification}

\begin{figure}[t]
\epsscale{0.9}
\plotone{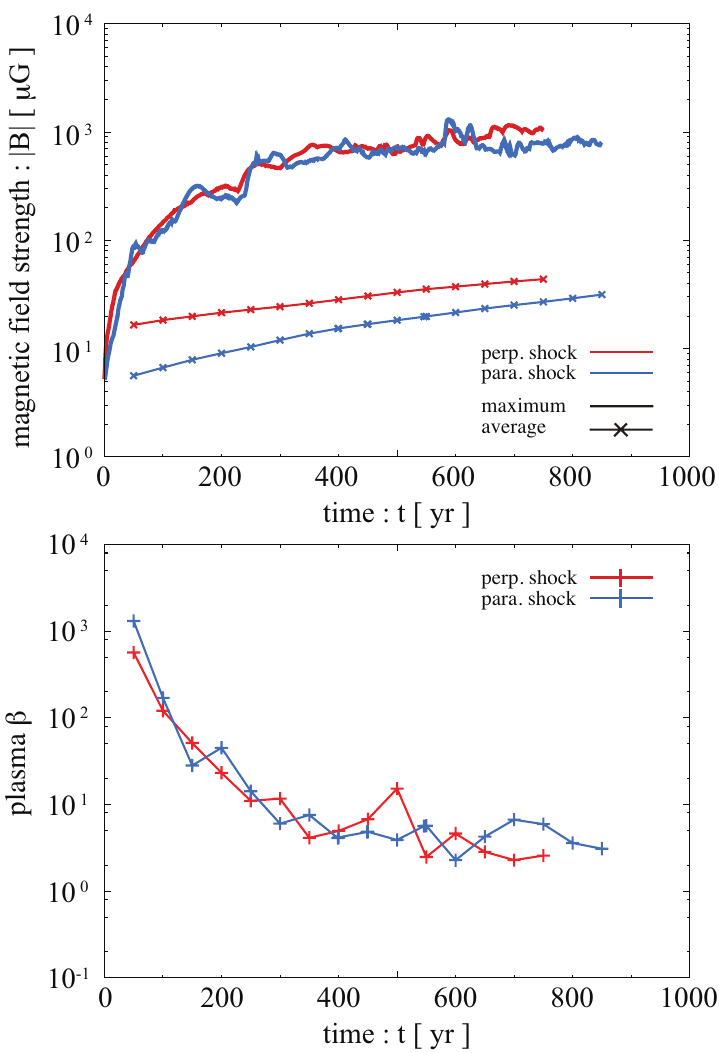}
\caption{
Top panel shows the evolution of maximum magnetic field strength and average field strength in the shocked gas.
Bottom panel shows evolution of plasma $\beta$ where the magnetic field strength is maximum.
Red and blue lines are the results of the perpendicular and the parallel shock cases, respectively.
}
\label{f5}
\end{figure}

In the post-shock region, the magnetic field is strongly amplified far beyond the shock compression value, which is very similar to the previous simulations (Inoue et al. 2009, 2010).
The top panel in Fig. \ref{f5} shows the plots of the evolutions of maximum magnetic field strength and average field strength $\langle |B| \rangle$ in the shocked slab.
Red and blue lines are results of the perpendicular and the parallel shock cases, respectively.
In both the perpendicular and parallel shock cases, the maximum magnetic field strength reaches on the order of 1 mG, which is 200 times larger than the initial strength.
We also plot the evolution of plasma $\beta \equiv 8\pi\,p/B^2$ where the magnetic field strength is maximum in the bottom panel of Fig. \ref{f5}.
It is clear that the saturation of the magnetic field amplification is determined by the condition of $\beta\sim1$.

One can understand the mechanism of this strong amplification using the following equation that is obtained from the equation of continuity and the induction equation:
\begin{equation}\label{IE}
\frac{d}{dt}\left( \frac{\vec{B}}{\rho} \right)=\frac{1}{\rho}\,(\vec{B}\cdot\vec{\nabla})\,\vec{v}.
\end{equation}
This equation indicates that the magnetic field is amplified, if the velocity has a shear along the field line.
We can also express the amplification as Faraday's law of induction, since the velocity shear along the magnetic field line is equivalent to a rotation of the electric field.
Such a situation is realized, in particular, at the transition layer between the cloud and diffuse gas due to the baroclinic effect.
From equation (\ref{IE}), we can estimate the timescale of the magnetic field amplification around the clouds as $t_{\rm grow}\sim l_{\rm tr}/\Delta v$, where $l_{\rm tr}$ is the thickness of the transition layer between the cloud and diffuse gas and $\Delta v$ is a velocity difference of the post-shock gas flows between the cloud and the diffuse gas (or a difference of the shock propagation speed).
Since the shock speed in the cloud is much smaller than that in the diffuse gas, the velocity difference $\Delta v$ can be estimated by the shock speed in the diffuse gas ($\sim 2000$ km s$^{-1}$).
The thickness of the transition layer is given by the so-called ``Field length'' $l_{\rm tr}\sim l_{\rm F}\equiv\sqrt{\kappa\,T/\rho\,{\cal L}}$, which expresses the thermal balance between structure formation by cooling and diffusion due to conduction (Field 1965, Begelman \& McKee 1990), where $\kappa$ is the thermal conductivity and ${\cal L}(n,T)$ is the cooling rate per unit mass.
In a typical ISM, $l_{\rm tr}$ takes a value of 0.05 pc (Inoue et al. 2006).
The above values give the growth timescale of $t_{\rm grow}\sim 20$ yr.
The results of simulations show that the growth times of the magnetic field at which the maximum magnetic field strength achieve $f_{\rm comp}\,B_{\rm ini}\,e$ (one $e$-folding time of the amplification after the shock passage, where $f_{\rm comp}=4$ for the perpendicular shock case and $f_{\rm comp}=1$ for the parallel shock case) are 39 yr for the perpendicular case and 17 yr for the parallel shock case, which are roughly in agreement with the above estimation.
Note that this strong amplification can be calculated correctly only if the scale of the transition layer $\sim 0.05$ pc is well resolved.
Our simulation with $\Delta x \simeq2\times 10^{-3}$ pc satisfactorily expresses the transition layer with more than 25 grid cells.
Also note that the strength of the vorticity generated behind the shock in the surface region of a cloud is determined by the upstream Field length, because the baroclinic vorticity is caused by the interaction of upstream density gradient and the shock.
Thus, the Field length mentioned above indicates the Field length in the upstream medium.

So far, we have only discussed the magnetic field amplification at the transition layer of the clouds and diffuse gas.
However, amplifications also arise in the diffuse gas, because the average magnetic field strength $\langle |B| \rangle$ shown in Fig. \ref{f5} mainly reflects the magnetic field strength in the diffuse gas.
In fact, in Fig. \ref{f2} and Fig. \ref{f3}, we can confirm the amplification in regions other than the thin transition layers.
As we have pointed out in \S \ref{Sturb}, turbulent eddies in the diffuse gas are induced by the effect of curved shock wave.
The induced eddies amplify the magnetic field through the stretching of magnetic field lines, or, in other words, through the action of the turbulent (or small-scale) dynamo effect (Brandenburg \& Subramanian 2005, Cho \& Vishniac 2000, Cho et al. 2009).
Note that such an amplification as the result of interaction between a shock and density fluctuation has been recognized by many authors (Giacalone \& Jokipii 2007,  Inoue et al. 2009, 2010, 2011, Mizuno et al. 2010).
Thus, as a consequence of the shock-cloud interaction, we can observe the two types of magnetic field amplification in response to the two mechanisms of vorticity generation.

Prior to the recognition of the magnetic field amplification by the interaction between a shock and density fluctuation, Balsara et al. (2001, 2004) found that the interaction between the shock and interstellar turbulence induce magnetic field amplification.
By measuring the field line stretching and the Lyapunov exponents of the chaotic, helical flows created by the shock-turbulence interaction, Balsara \& Kim (2005) showed that stretch, twist and fold (STF; Childress \& Gilbert 1995) of magnetic field line plays a crucial role in the magnetic field amplification.
On the other hand, in the interaction between a shock and density fluctuation, we have somewhat strong evidence that shows the STF is not playing a main role in the amplification, because the result of the previous 2D simulations with almost the same initial condition, in which the STF is definitively unable to work, shows very similar magnetic field amplification (Inoue, Yamazaki, \& Inutsuka 2009, see also similar 2D simulations by Giacalone \& Jokipii 2007 and Mizuno et al. 2010).
This indicates that the magnetic energy in the results of our simulations is not mainly contained in the mean-field due to the STF, but contained in waves and/or locally stretched magnetic field lines generated by vortical and shear flows.
Note that this result does not contradict Balsara's conclusion, because the simulations by Balsara et al. (2001, 2004) involves realistic turbulence in upstream ISM, while the simulations of the density fluctuation and shock interaction do not.
We expect that if our initial medium involved turbulence that creates large helicity density by the shock-turbulence interaction as shown in Balsara et al. (2001, 2004), we might also obtain the additional mean-field amplification by the STF.
This speculation will be tested in our future study of the interaction between turbulent molecular cloud and a supernova blast wave.

\begin{figure}[t]
\epsscale{1.}
\plotone{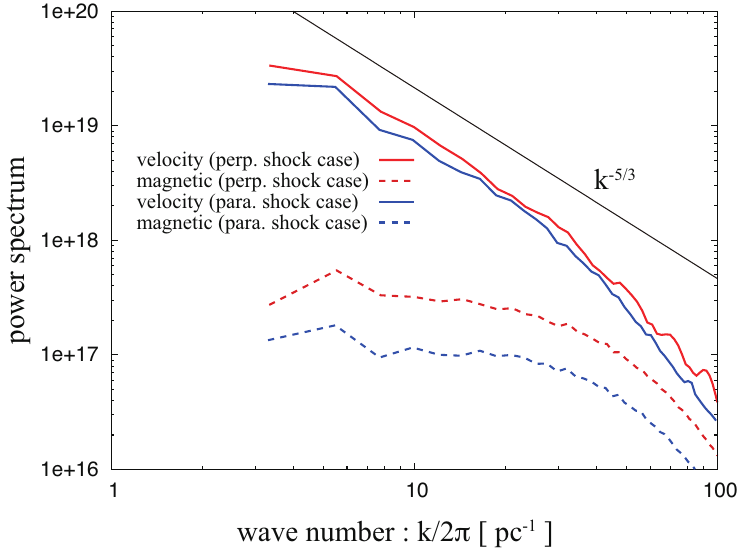}
\caption{
Power spectra of velocity ({\it solid}) and magnetic field ({\it dashed}).
Red and blue lines represent the perpendicular and the parallel shock cases, respectively.
Spectra are calculated using the data of the turbulent regions of $x\in [1.0,1.5 ]\,\cap\,y\in [1.3,1.8 ]\,\cap\,z\in [0.0,0.5 ]$ (for the perpendicular shock case) and $x\in [0.3,0.8 ]\,\cap\,y\in [1.1,1.6 ]\,\cap\,z\in [0.9,1.4 ]$ (for the parallel shock case) at $t=750$ yr.
The black line shows the Kolmogorov law $k^{-5/3}$.
}
\label{f6}
\end{figure}

\subsection{Spectra of Turbulence}
Fig. \ref{f6} shows the one-dimensional power spectra of the velocity ($P_v(k)$; {\it solid}) and magnetic field ($P_B(k)$; {\it dashed}), where the power spectra are defined as $\int P_v\,dk=\int v^2\,d^3x$ and $\int P_B\,dk=\int B^2\,d^3x$.
Red and blue lines represent the perpendicular and the parallel shock cases, respectively.
Spectra are calculated using the data of the turbulent regions of $x\in [1.0,1.5 ]\,\cap\,y\in [1.3,1.8 ]\,\cap\,z\in [0.0,0.5 ]$ for the perpendicular shock case and $x\in [0.3,0.8 ]\,\cap\,y\in [1.1,1.6 ]\,\cap\,z\in [0.9,1.4 ]$ for the parallel shock case at $t=750$ yr.
These regions are selected so that more than 85\% of the regions are filled by the shocked gas.
The obtained spectra are very similar to those of the super-Alfv\'enic turbulence with isotropic, large-scale, divergence-free forcing (Cho \& Vishniac 2000, Cho \& Lazarian 2003, Cho et al. 2009).
In large scales ($l\equiv2\pi/k\gtrsim0.03$ pc), the velocity field shows the Kolmogorov spectrum $\propto k^{-5/3}$, and the spectrum of the magnetic field is nearly flat.
In small scales, the power spectra go down likely due to numerical dissipation.

Theoretically, below the scale where the velocity dispersion becomes comparable to the mean field Alfv\'en velocity due to the cascade of the turbulence, the cascade mechanism can be changed from the Kolmogorov-like one to the MHD critical balance cascade (Goldreich \& Sridhar 1995).
Since the velocity dispersion of Kolmogorov turbulence depends on the scale as $l^{1/3}$, if we use the driving scale of turbulence $L\sim 0.2$ pc from Fig. \ref{f6}, the velocity dispersion at the driving scale $\langle v \rangle_{L}\sim 700$ km/s, and the mean field Alfv\'en velocity $\langle v_{\rm A} \rangle \sim 100\mbox{ km s}^{-1}\,(|\langle \vec{B}\rangle|/100\,\mu\mbox{G})\,(\langle n\rangle/4\,\mbox{cm}^{-3})^{1/2}$, the transition scale of the cascade mechanism can be evaluated as $l_{\rm cas}\sim L \langle v_{\rm A} \rangle^3/\langle v \rangle_{L}^3 \simeq 5\times10^{-4}$ pc.

In the scales larger than $l_{\rm cas}$, we can expect large-amplitude isotropic magnetic field fluctuations, because the magnetic field is passive with respect to the turbulent velocity field.
Thus, the resonant scattering of particles by Alfv\'en waves would be very effective and we can expect the Bohm-limit diffusion ($\delta B^2/B^2\sim1$), if the resonant scale of particles $l_{\rm res}=2\pi\,E/e\,B\simeq 7\times10^{-3}\,\mbox{pc}\,(E/100\,\mbox{TeV})\,(B/100\,\mu\mbox{G})^{-1}$ is larger than $l_{\rm cas}$.
In the scales smaller than $l_{\rm cas}$, according to Yan \& Lazarian (2002), the resonant scattering of particles by Alfv\'en waves would be ineffective due to the anisotropy of MHD turbulence.
However, this does not indicate the inefficiency of the scattering by the MHD turbulence for the particles whose resonant scale is smaller than $l_{\rm cas}$.
Beresnyak et al. (2010) recently studied the transport of test particles in a compressible MHD turbulence.
They found that the particles are effectively scattered by non-resonant mirror reflections due to the large-scale magnetic field fluctuations.
Thus, the MHD turbulence induced in the simulations would be important for the particle scattering even for the low-energy particles whose the resonant scale is smaller than $l_{\rm cas}$.

\begin{figure}[t]
\epsscale{0.9}
\plotone{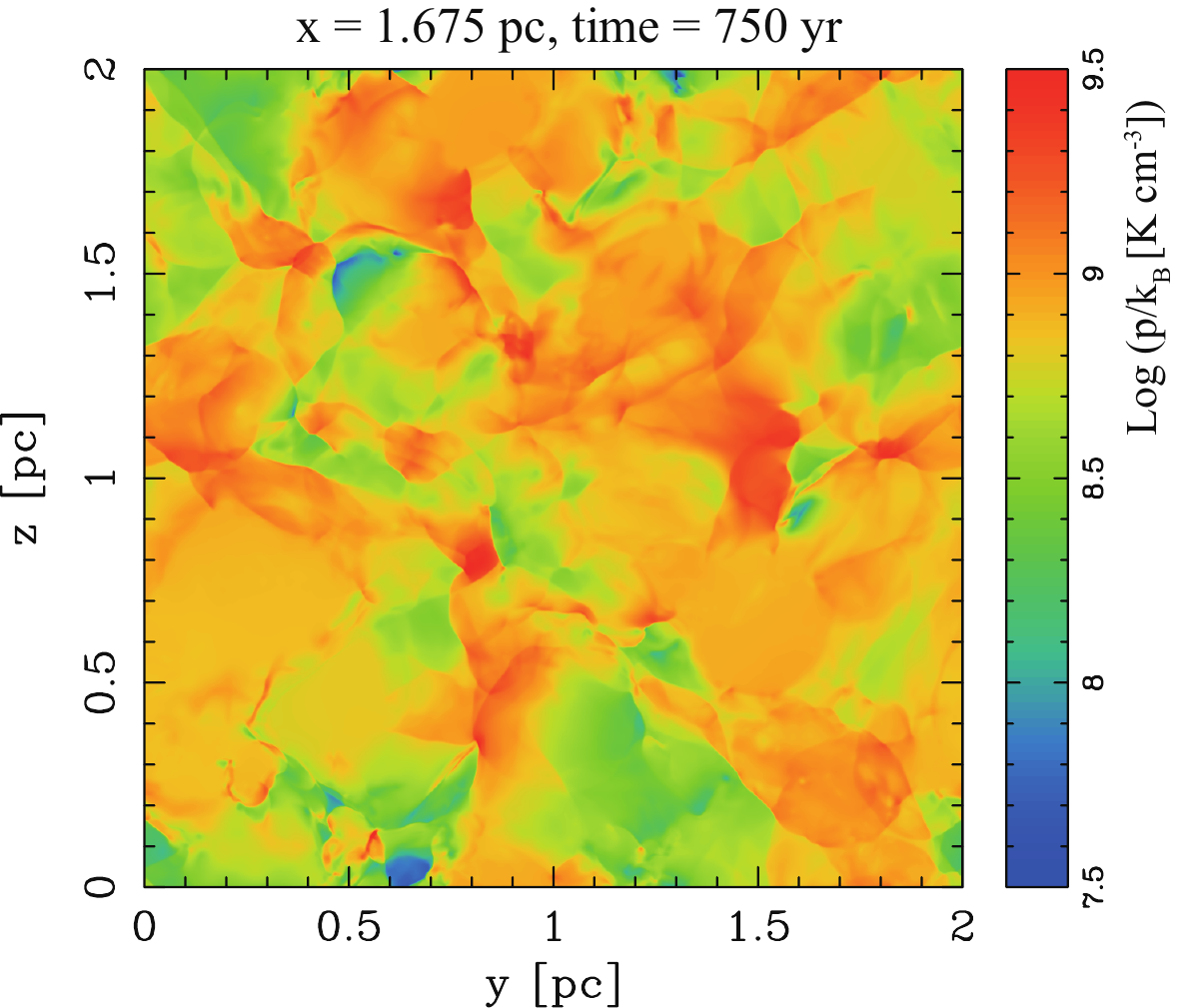}
\caption{
Pressure slice at the $x=1.675$ pc plane of the perpendicular shock case at $t=750$ yr.
}
\label{f7}
\end{figure}

\subsection{Transmitted and Reflected Shock Waves}

When the strong shock wave hits a cloud, a transmitted shock wave penetrates into the cloud and a reflected shock wave propagates back into the shocked slab.
Since the reflected shocks are formed at the surface of the clouds, many reflected shock waves are generated in the SNR.
For instance, in Fig. \ref{f7}, we show the pressure slice of the perpendicular shock case in the $y$-$z$ plane at $x=1.675$ pc.
We can see a number of discontinuous pressure jumps due to the reflected shocks in the SNR.
In the following, we also describe the reflected shocks as the secondary shocks to distinguish them from the primary forward shock wave.

In Appendix A, we analytically evaluate the Mach number of the reflected shock using a formulation given by Miesch \& Zweibel (1994).
We find that the Mach number of the reflected shock wave is a function of the density ratio of the intercloud gas and cloud (see eq. [\ref{tr3}]).
In the case of our simulation, the ratio $n_{\rm i}/n_{\rm c}\sim40$ gives the Mach number of the secondary shocks as $M\simeq 1.8$.
We also show in Appendix A that the Mach numbers of the shock wave in diffuse gas and the transmitted shock wave in clouds are almost identical.

\begin{figure}[t]
\epsscale{1.}
\plotone{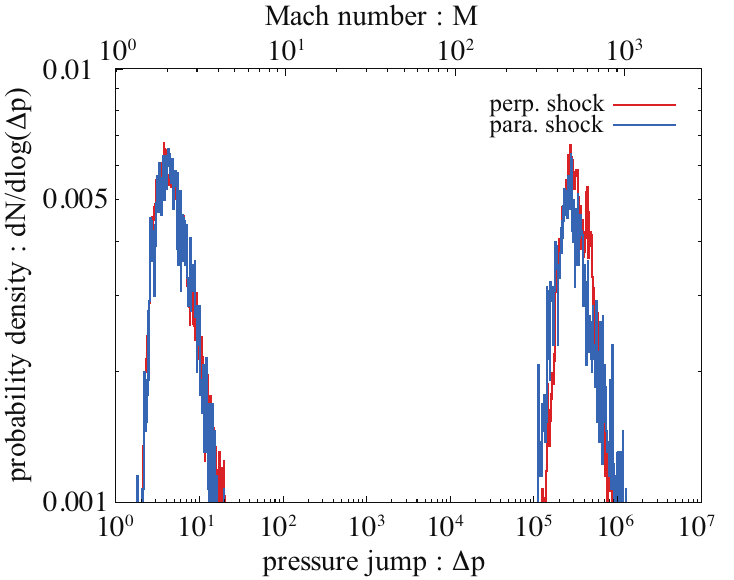}
\caption{
Probability distribution function of pressure jump $\Delta p$.
Red and blue lines are the results of the perpendicular and the parallel shock cases, respectively.
Upper horizontal axis denote the Mach number of shock waves converted by using eq. (\ref{RH}). 
}
\label{f8}
\end{figure}

To evaluate the strength of the transmitted and reflected shocks in the simulations, we plot the probability distribution function (PDF) of the pressure jumps in Fig. \ref{f8}.
The PDF is calculated using the following procedure:
We define the sphere whose radius is 20 times the numerical grid size and whose center is at each cell in the numerical domain.
In each sphere, we calculate the minimum pressure $p_{\rm min}$, the maximum pressure $p_{\rm max}$, and the maximum pressure gradient, except for the injected hot plasma with $n<0.5$ cm$^{-3}$. 
If the maximum pressure gradient is larger than the critical pressure gradient $p_{\rm cr}=p_{\rm max}/(5\,\Delta x)$, i.e., the pressure fluctuation in the sphere is caused within the narrow region, we regard a shock wave as being in the sphere whose pressure jump is $\Delta p=p_{\rm max}/p_{\rm min}$.
Here the factor 5 in the expression of the critical pressure gradient is chosen from the fact that the high-resolution shock-capturing scheme used in this study requires 5 grid points at most to express a shock wave.
The bimodal distribution of the pressure jump in Fig. \ref{f8} shows the existence of the primary shock and secondary shocks.
According to the Rankine-Hugoniot relation, the Mach number of a shock for the gas of $\gamma=5/3$ is related to the pressure jump as (Landau \& Lifshitz 1959):
\begin{equation}\label{RH}
{\cal M}(\Delta p)=\sqrt{ \frac{1}{5}+\frac{4}{5}\,\Delta p}.
\end{equation}
Substituting the peak of the PDF $\Delta p=3.85$ that corresponds to the secondary shocks, we obtain ${\cal M}=1.81$, which agrees well with the above-mentioned analytical evaluation ${\cal M}\simeq1.8$ (see, Appendix A for more detail).
In the upper horizontal axis of Fig. \ref{f8}, we also exhibit the Mach number of shock waves converted by using eq. (\ref{RH}). 

The peaks of the PDF with higher pressure jumps are composed of the pressure jumps due to the primary shock propagating in the diffuse gas and the transmitted shocks in the clouds.
This also agrees well with the analytical evaluation that the Mach numbers of both the shock in the diffuse gas and the transmitted shock are almost identical, i.e., the primary shock propagates so that the Mach number remains unchanged even after the encounter with the cloud.

Note that even in the case of a different initial medium, e.g., a medium with larger clump density and larger clump filling factor, the Mach number of the secondary shocks would be affected marginally.
This is because the Mach number of the reflected shock depends only weakly on the density ratio of the intercloud gas and cloud, and the Mach number converges to ${\cal M}=\sqrt{5}$ in the limit of a large density ratio (see Appendix A).

\section{Discussion: Observational Features}
\subsection{Short-Time X-ray Variability}
From the observations of RX J1713.7$-$3946 using the {\it Chandra} and {\it Suzaku X-ray space telescopes}, Uchiyama et al. (2007) discovered that the synchrotron emissions from X-ray hot spots whose spatial scale is $\sim0.05$ pc show timescale variability of a few years.
They concluded that the short-time variability would be a consequence of magnetic field amplification up to 1mG.
This is because the timescale of the synchrotron cooling that causes the decay of X-ray luminosity, and the acceleration timescale of the Bohm-limit DSA that causes brightening, are respectively given by (Uchiyama et al. 2007, Malkov \& Drury 2001)
\begin{equation}\label{xhs1}
t_{\rm syn}\simeq1.5\,\left( \frac{B}{1\,\rm{mG}} \right)^{-1.5}\,\left( \frac{\epsilon}{1\,\rm{keV}} \right)^{-0.5}\,\rm{yr},
\end{equation}
\begin{equation}\label{xhs2}
t_{\rm acc}\simeq1\,\eta\,\left( \frac{B}{1\,\rm{mG}} \right)^{-1.5}\,\left( \frac{\epsilon}{1\,\rm{keV}} \right)^{0.5}\,\left( \frac{v_{\rm sh}}{3,000\,\rm{km\,s}^{-1}} \right)^{-2}\,\rm{yr},
\end{equation}
where $\eta=\delta B^2/B^2$ is the degree of magnetic field fluctuations that characterizes the efficiency of the acceleration and $\epsilon=1.6\,(B/1\mbox{mG})\,(E/10\mbox{TeV})^2$ keV is the energy of synchrotron photons emitted from accelerated electrons with energy $E$.

The results of our simulations can explain such a short-time variability of the synchrotron X-rays.
As shown in \S 3.2, the strong magnetic field amplification easily makes regions with $B\sim1$ mG, and their spatial scale $\sim 0.05$ pc also agrees quite well with the observed regions.
Since our results show the regions of $B\sim 1$ mG are downstream of the primary shock wave, an additional accelerator other than the primary shock is necessary to produce brightening of the X-rays.
The secondary shocks generated by the shock-cloud interaction can be such accelerators.
Although the Mach number of the secondary shocks is much smaller than that of the primary shock, their propagation speed is comparable to the primary shock's, because they are downstream of the primary shock where the sound speed is comparable to the primary shock speed in a diffuse medium.
Thus, $v_{sh}\sim 3000$ km s$^{-1}$ in eq. (\ref{xhs2}) is available even when the accelerator is the secondary shock.
Owing to the small Mach number of the secondary shocks, the injection rate of the electron acceleration may be much smaller than the case of the primary shock.
Even so, the secondary shocks can accelerate electrons, because the secondary shocks are in the downstream region of the primary shock where the relativistic electrons accelerated by and advected from the primary shock are available as seed particles for further acceleration.
Note that, in order to adopt the estimation of eq. (\ref{xhs2}), both upstream and downstream magnetic field should be amplified and such high-field region should be larger than the acceleration sites, i.e., the region in which the particles being accelerated exist.
Again, since the secondary shocks propagate in the shocked region, the magnetic field is already amplified upstream of the secondary shocks and electrons with an energy of 10 TeV may go upstream and reach the distance of $\eta\,l_{g}\,c/v_{sh}\simeq 5\times 10^{-3}\,\eta$ pc from the shock front (assuming $B\sim1$ mG), which is smaller than the scale of the milli-Gauss regions $\sim 0.05$ pc.
Therefore, the acceleration timescale of a few years is basically possible at the secondary shocks.
Note that the spectrum of particles reaccelerated by the secondary shock would be steeper than that accelerated by the primary shock because of the small Mach number of the secondary shocks.
This indicates that the synchrotron emission of electrons reaccelerated by the secondary shocks can dominate the emission of electrons accelerated by the primary shock only in the regions with high magnetic field strength (i.e., the electrons accelerated by the primary shock are already cooled)\footnote{In the case of a middle-aged SNR, since the acceleration by the primary shock can be ineffective above the break energy of $\sim 10$ GeV due to the damping of MHD waves by ion-neutral collisions, the particles accelerated by the secondary shocks can dominate the spectrum above the break energy (see, Inoue, Yamazaki, \& Inutsuka 2010 for detail). This is a possible origin of the broken power-law spectrum of particles commonly observed in middle-aged SNRs interacting with molecular clouds.}.

If we only take into account the resonant scattering of particles by turbulent MHD waves, the Bohm-limit acceleration ($\eta=1$) in eq. (\ref{xhs2}) may not be achieved because the resonant scale for relativistic electrons $l_{\rm res}=2\pi\,E/e\,B\simeq 7\times10^{-5}\,(E/10\mbox{TeV})\,(B/1\mbox{mG})^{-1}$ pc is somewhat smaller than the transition scale of the cascade mechanism ($l_{\rm cas}\sim 5\times 10^{-4}$) above which the Bohm-limit diffusion can be expected (see \S 3.3).
However, as discussed in \S 3.3, we can additionally expect the non-resonant mirror scatterings by the large-scale magnetic fluctuations that would make the Bohm-limit acceleration possible in the turbulent shell.
Furthermore, according to Ohira et al. (2009), the effect of a cold proton beam created by a charge-exchange process in shocked neutral gas causes large-amplitude magnetic field fluctuations through the growth of the Weibel instability or the resonant instability that also enhances the acceleration efficiency around the shocked clouds.
It is worth noting again that our scenario explains not only the timescale of X-ray variability but also its typical spatial scale of $\sim 0.05$ pc, since the strong magnetic field amplifications to milli-Gauss take place at the transition layers between the clouds and diffuse gas that is essentially determined by the Field length $\sim 0.05$ pc (Inoue et al. 2006).

The evidence of magnetic field amplification in RX J1713.7$-$3946 has also been obtained from the thickness of X-ray filaments that typically show $B\sim 100$ $\mu$G (Hiraga et al. 2005, Ballet 2006).
Since the X-ray filaments have apparently coherent features on parsec scales, it seems to be the average field strength.
The average field strengths in our simulations are tens of micro-Gauss (see Fig. \ref{f5}), which possibly explains the X-ray filaments.
However, the average strength in our simulations becomes noticeable only in the downstream region far from the primary shock front suggesting that other possible amplification mechanisms such as the cosmic-ray streaming instability (Lucek \& Bell 2000) and/or dynamo effect driven by the interaction between cosmic-ray precursor and density fluctuation (Beresnyak et al. 2009) might be important in the vicinity of the primary shock front.
We also stress that as pointed out by Pohl et al. (2005), the evaluation of the magnetic field strength in the X-ray filaments based on the synchrotron cooling timescale might be questionable, i.e., the thickness of the filaments might be determined by the spatial scale of the strong magnetic-field region instead of the cooling length of electrons in a uniformly magnetized region.

\subsection{Global Morphology of SNR}
One may wonder that despite its associating molecular clouds, especially in the northeast region, the morphology of RX J1713.7$-$3946 is roughly spherical and the expansion rate is fast as if it is evolving in the diffuse environment.
This is not surprising if we take into account the clumpy nature of the molecular clouds:
As we showed in \S3.1, the mean propagation speed of the primary shock is determined by that in the diffuse gas.
The speed of the primary shock is stalled locally where it hits a clump of the clouds.
However, once the shock passes the clump or once the shock entirely surrounds the clump and is topologically separated from the clump, the shock is accelerated and the deformation is dissolved, because the fast shock is stable with respect to the deformation.
Thus, the local configuration of the shock wave can be deformed only where the shock is interacting with local clumps, while the global configuration would be affected only if the density of the diffuse surrounding of the clouds has global variations.

\subsection{Suppression of Thermal X-ray Line Emission}
It has been shown that RX J1713.7$-$3946 is characterized by the lack of thermal X-ray emission (Slane et al. 1999, Cassam-Chenai et al. 2004, Takahashi et al. 2008).
According to Takahashi et al. (2008), the upper limit of gas density to ensure the thermal X-ray-less feature is 0.01 to 2 cm$^{-3}$ depending on the assumed electron temperature.
The upperlimit density ($n\sim0.01$-$2$ cm$^{-3}$) is smaller than the mean density of the diffuse ISM ($n\sim 1$ cm$^{-3}$), if the electron temperature is larger than 0.05 keV.
In addition, if the temperature of shocked clouds is larger than $\sim1$ keV, they would emit an unacceptable amount of thermal X-ray lines because of their high density.
However, if we take into account the effect of stellar wind from the progenitor of a supernova, the density of the diffuse gas can be smaller than the threshold density for thermal X-ray line emission.
Furthermore, the thermal X-ray line emission from the shocked clouds are substantially suppressed because of the stall of the transmitted shock.
These two effects on the thermal X-ray line emission were pointed out in Zirakashvili \& Aharonian (2010).
In the following, we discuss these effects more quantitatively based on the results of our simulations.

\begin{figure}[t]
\epsscale{1.}
\plotone{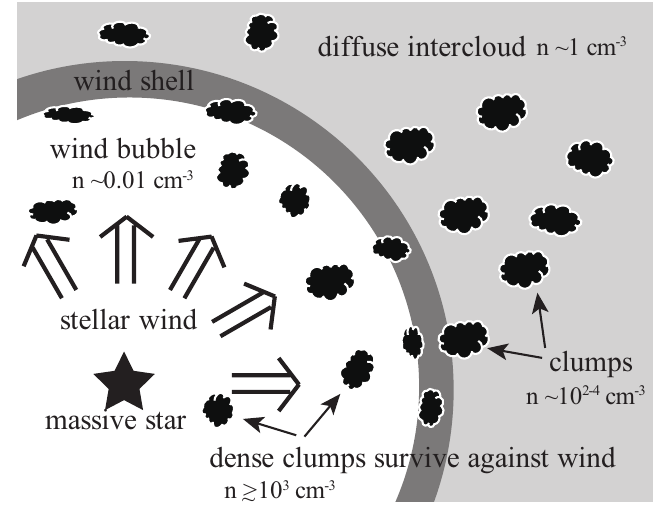}
\caption{
Schematic view of wind bubble expanding in a cloudy ISM.
Diffuse intercloud gas is swept by the stellar wind, while dense cloud cores and clumps can survive in the wind.
Density in the wind bubble is much smaller than the intercloud gas density that is determined by the evaporation of the wind shell by thermal conduction.
 }
\label{f9}
\end{figure}

The requirement for the density of the diffuse gas can be achieved if the progenitor of RX J1713.7$-$3946 is a massive star as is widely believed (Slane et al. 1999).
This is because the stellar wind from the massive star would sweep up pre-existing intercloud gas rarefying the intercloud gas significantly, while dense clumps are not swept off owing to their high density (e.g., Gritschneder et al. 2009).
The situation is illustrated schematically in Fig. \ref{f9}.
According to Weaver et al. (1977), who studied the expansion of a bubble formed by stellar wind from O-type stars, the resulting gas density in the wind bubble is $n\sim 0.01$ cm$^{-3}$ (see, e.g., Fig. 3 of Weaver et al. 1977).
Note that the density in the wind bubble is not determined by the density of wind gas but by the evaporation of the wind shell into the cavity.
The radius of stellar wind bubble $R_{\rm w}$ is described using the mechanical luminosity of the wind $L_{\rm w}$, density of interstellar gas $n_0$, and lifetime of the wind $t_{\rm life}$ as $R_{w}=27$ pc $(L_{\rm w}/10^{36}$ erg s$^{-1})^{1/5}\,(n_{0}/1$ cm$^{-3})^{-1/5}\,(t_{\rm life}/1$ Myr$)^{3/5}$ (Castor et al. 1975).
According to this expansion law, in order for the dense gas to stay within the cavity of the wind bubble, the density should be at least larger than
\begin{equation}\label{nth}
n_0\gtrsim 10^3 \mbox{ cm}^{-3} \left( \frac{L_{\rm w}}{10^{36}\mbox{ erg s}^{-1}} \right) \left( \frac{R_{\rm w}}{10\mbox{ pc}} \right)^{-5} \left( \frac{t_{\rm life}}{1\mbox{ Myr}} \right)^{3},
\end{equation}
where we have adopted a distance of 1 kpc and thus the radius of RX J1713.7$-$3946 $\sim 10$ pc (Fukui et al. 2003).

Recently, Sano et al. (2010) have shown by using the {\it NANTEN telescope} that the ``peak C'' of a CO molecular cloud core associated with the region in RX J1713.7$-$3946 seems to be embedded in the SNR.
Since the density of the molecular cloud core is approximately $10^4$ cm$^{-3}$, it is reasonable for such a dense object to stay in the SNR.
Eq. (\ref{nth}) suggests that less dense molecular cloud cores or molecular clumps with density on the order of $10^3$ cm$^{-3}$ depending on $L_{\rm w}$ and $t_{\rm age}$ would also be embedded in RX J1713.7$-$3946, although these may not be observed by CO line emission surveys due to the dissociation of molecules by UV radiations from the progenitor massive star.
We conclude that if we take into account the effect of the stellar wind from the massive progenitor, the diffuse intercloud gas density becomes on the order of $n\sim 0.01$ cm$^{-3}$, which does not conflict the lack of the thermal X-ray line emission, while dense molecular clumps/cores can be left in the wind bubble.

The remaining issue for the X-ray line emission from the shocked clouds is resolved easily as follows:
The temperature of protons in the shocked gas, which corresponds to the maximum temperature of electrons, is given by
\begin{equation}
k_{\rm B}\,T=\frac{3}{16}\,m_{\rm p}\,v_{\rm sh}^2=18\,\left( \frac{v_{\rm sh}}{3,000\,{\rm km}\,{\rm s}^{-1}} \right)^2\,{\rm keV},
\end{equation}
where $v_{\rm sh}$ is the shock velocity that is supposed to be 3000 km s$^{-1}$ in the diffuse gas (gas in the wind cavity with the density $n_{\rm d}\sim 0.01$ cm$^{-3}$).
In the cloudy ISM, the shock is stalled when it hits a cloud.
As we show in \S 3.1 and Appendix A in more detail, the velocity ratio of the shock wave in the diffuse gas and the cloud is proportional to the square root of their density ratio: $v_{\rm sh,d}/v_{\rm sh,c}\simeq (n_{\rm c}/n_{\rm d})^{1/2}$.
From this relation, we can estimate the proton temperature (corresponding to the upper bound of the electron temperature) of the shocked cloud as
\begin{eqnarray}
k_{\rm B}\,T_{\rm c}&=&\frac{3}{16}\,m_{\rm p}\,v_{\rm sh,c}^2 \nonumber \\
&=&2\times10^{-4}\,\left( \frac{v_{{\rm sh},d}}{3,000\,{\rm km}\,{\rm s}^{-1}} \right)^2 \, \left( \frac{n_{\rm d}}{0.01\,{\rm cm}^{-3}} \right) \nonumber \\
&&\times \left( \frac{n_{\rm c}}{10^3\,{\rm cm}^{-3}} \right)^{-1}\,{\rm keV}. \label{vsh0}
\end{eqnarray}
Therefore, even after the passage of the shock wave in the clouds, bright thermal X-ray line emission from the clouds is not expected.

\subsection{Spectrum of Hadronic Gamma Rays}
Recently, using one-dimensional model assuming a uniform ISM, Ellison et al. (2010) claimed that if we reduce the ambient density to reconcile the absence of the thermal X-ray line emission from RX J1713.7$-$3946, the hadronic gamma-ray emission becomes dim owing to the low target-gas density for $\pi^0$ creation.
The reason is as follows:
According to Aharonian et al. (2006), the total gamma-ray energy measured from 0.2 to 40 TeV in RX J1713.7$-$3946 is $W\simeq 6\times10^{49}\,(d/1\mbox{ kpc})^2\,(n_{\rm tg}/1\mbox{ cm}^{-3})^{-1}$ erg, where $d$ is a distance and $n_{\rm tg}$ is a mean target gas density.
Thus, supposing the low-density ISM, the efficiency of particle acceleration becomes $100\,(n_{\rm tg}/0.06\mbox{ cm}^{-3})\,(E/10^{51}\mbox{ erg})\%$ indicating that the hadronic gamma-ray emission cannot be as bright as observed even if the acceleration is extremely efficient.

However, in our shock-cloud interaction model, the hadronic emission from the clouds embedded in the SNR can be expected, because the high density shocked clouds do not emit thermal X-ray lines owing to the low-temperature as shown in eq. (\ref{vsh0}).
If we assume a typical density of clumps $n_{\rm cl}\sim 10^{3}$ cm$^{-3}$ and their volume filling factor $f\sim 10^{-3}$, the effective mean target density can be rewritten as $n_{\rm tg}\simeq n_{\rm cl}\,f$ and thus the efficiency becomes $6\,(n_{\rm cl}/10^{3}\mbox{ cm}^{-3})\,(f/10^{-3})\%$.
Although precise evaluation of the filling factor $f$ is beyond the scope of this paper, our model can reproduce the hadronic gamma-ray emission that is compatible with the canonical acceleration efficiency $\sim 10$\%.

In the case of a uniform ISM model, the spectral energy distribution of the hadronic gamma rays directly reflects that of the accelerated nuclei roughly above the critical energy of the $\pi^0$ creation ($\sim0.1$ GeV), i.e., the photon index of the hadronic gamma-ray emission is $p=2$ for the standard DSA scenario.
However, in our shock-cloud interaction model, as we discuss in the following, the spectrum may deviate from this conventional spectrum.
Because of the heavy shock stall in the clouds, particle acceleration at the transmitted shock wave would be inefficient and most of the particles could be accelerated at the primary shock in the diffuse gas.
The high-energy nuclei accelerated at the primary shock interpenetrate diffusively to nearby shocked dense clumps.
The amount of the neutral pions generated through the collisions of accelerated nuclei and matters in the clouds, which eventually decay and emit gamma rays (Issa \& Wolfendale 1981, Naito \& Takahara 1994, Aharonian et al. 1994), is proportional to the mass of the dense clumps illuminated by the high-energy nuclei.
Since the penetration depth into the clouds would depend on the energy of accelerated particles, the hadronic gamma-ray spectrum can be deviated from that of the conventional model supposing uniform ISM (see also Zirakashvili \& Aharonian 2010). 

In our simulations, the clouds formed by the thermal instability have a sheet-like structure.
Using the penetration depth $l_{\rm pd}$ of the accelerated protons into the cloud, the mass of the cloud illuminated by the high-energy protons is $M\sim R^2\,l_{\rm pd}\,\rho$, where $\rho\simeq 40$ cm$^{-3}$ is the density of clouds and $R$ is the scale of the cloud sheet $\sim 1$ pc.
This indicates that the mass of the interpenetrated region of the cloud is proportional to the penetration depth as long as the penetration depth is smaller than the thickness of the cloud sheet $\simeq 0.1$ pc.
However, as discussed in \S 4.3, low-density cloud envelope would be wiped out by the stellar wind.
In the case of RX J1713.7$-$3946, the clouds that remain in the SNR would be dense molecular cloud cores/clumps as suggested in eq. (\ref{nth}) and observationally in Sano et al. (2010).
If the dense molecular cloud cores are gravitationally bound, collapsing objects, the density structure becomes $\rho(r)=\rho_0\,(r/r_0)^{-2}$ (Larson 1969, Penston 1969, Masunaga \& Inutsuka 2000, Andre et al. 2000).
This is indeed the case of the ``peak C'' of CO core embedded in RX J1713.7$-$3946 (Sano et al. 2010).
In this case the mass illuminated by the high-energy protons can be written as
\begin{equation}
M=\int_{R-l_{\rm pd}}^{R} 4\pi\,r^2\,\rho_0\,(r/r_0)^{-2}\,dr=4\pi\,\rho_0\,r_0^2\,l_{\rm pd}
\end{equation}
If the cores are gravitationally unbound, pressure-confined objects, the density would be approximately constant (Bonner 1956, Ebert 1955, Alves et al. 2001), and the illuminated mass can be written as
\begin{eqnarray}
M&=&\int_{R-l_{\rm pd}}^{R} 4\pi\,r^2\,\rho\,dr=\frac{4\pi\,\rho}{3}\,(3\,R^2\,l_{\rm pd}-3\,R\,l_{\rm pd}^2+l_{\rm pd}^3) \nonumber \\
&\simeq&4\pi\,\rho\,R^2\,l_{\rm pd}\,\,\,\mbox{for }R\gg l_{\rm pd},
\end{eqnarray}
The above examples indicate $M\propto l_{\rm pd}$ as long as $l_{\rm pd}\ll R$.
The radius of the molecular cloud cores is typically 0.1-1 pc, and the scale of dense clumps in molecular clouds is usually larger than $0.1$ pc because 0.1 pc is the minimum scale of the thermal instability (Field 1965, Koyama \& Inutsuka 2004, Inoue et al. 2006) and the turbulent flows that form clumps by shock compressions  is expected to be subsonic below 0.1 pc (Larson 1981, Heyer \& Brunt 2004, Mac Low \& Klessen 2004).
Note that when the density of clouds are as high as $10^3$ cm$^{-3}$, the cooling timescale of clouds becomes much smaller than the age of RXJ1713.7$-$3946 $\sim$1,000 years that substantially diminish the scale of shocked region of clouds due to cooling contraction.
However, owing to the slowdown of shock speed, the penetration length of the shock wave into cloud in 1,000 year is evaluated to be $\sim$0.01 pc (where we have assumed the density ratio between the clouds and diffuse gas to be $10^5$ as we discussed in \S 4.3) that is much smaller than the scale of the molecular cloud cores $\lesssim$ 0.1 pc.
Thus, we can still assume the minimum scale of clouds to be 0.1 pc, even if clouds are apparently embedded in SNR.

Since the diffusion coefficient for the high-energy particles can be written as $\kappa_{\rm d}=4\,\eta\,l_{\rm g}\,c/3\pi$ (Skilling 1975), where $l_{\rm g}$ is the gyro radius of relativistic particles, the penetration depth due to the random walk of the particles can be written as
\begin{eqnarray}
l_{\rm pd}&\simeq& (\kappa_{\rm d}\,t)^{1/2} \nonumber \\
&=& 0.1\,\eta^{1/2} \left( \frac{E}{10\mbox{ TeV}} \right)^{1/2} \left( \frac{B}{100\,\,\mu\mbox{G}} \right)^{-1/2} \left( \frac{t_{\rm age}}{10^3\mbox{ yr}} \right)^{1/2} \mbox{ pc},
\end{eqnarray}
where $E$ is the particle energy, and $t_{\rm age}$ is the age of the SNR.
For the magnetic field strength in the above expression, we have used the observed typical strength in the dense regions of molecular clouds with density $\sim 10^3$ to $10^4$ cm$^{-3}$ (Crutcher 1999).
The parameter $\eta=B^2/\delta B^2$ has large ambiguity.
As we have discussed in \S 3.3 and \S 4.3, the Bohm-limit diffusion can be expected not only around the shocked clouds but also in shocked clouds through the scattering by turbulent magnetic field fluctuations (Beresnyak et al. 2011) and fluctuations generated by the effect of a cold ion beam (Ohira et al. 2009).
Also, the observations by Uchiyama et al. (2007) suggest that $\eta\sim1$ is realized at least around the clouds.
Thus, if we assume $\eta\sim1$, the penetration depth for particles with $E\lesssim 10$ TeV can be smaller than the minimum scale of clouds $\sim 0.1$ pc and the interpenetrated mass can be proportional to the square root of particle energy $M(E)\propto l_{\rm pd}(E)\propto E^{1/2}$.

In the conventional one-zone model that assumes the amount of target matter creating $\pi^0$ is independent of the energy of accelerated protons, the spectral energy distribution of gamma rays becomes $N(E)\,dE\propto E^{-p}\,dE$ above the critical energy for the $\pi^0$ creation and below the maximum energy of accelerated nuclei.
Here $p$ is the spectral index of the distribution of high-energy nuclei with $p=2$ in the conventional DSA theory.
While in the shock-cloud interaction model, the amount of matter creating $\pi^0$ depends on the square root of the energy of accelerated protons, so that the gamma-ray distribution becomes $N_{\gamma}(E)\,dE\propto M(E)\,E^{-p}\,dE\propto l_{\rm pd}(E)\,E^{-p}\,dE\propto E^{-p+1/2}\,dE$.
This yields $N_{\gamma}(E)\propto E^{-1.5}$ for $p=2$: that is consistent with the recent observation of RX J1713.7$-$3946 (Abdo et al. 2011).
This photon index of the hadronic gamma-ray emission ($p-1/2$) is the same as that of the inverse Compton emission $(p+1)/2$ when $p=2$.
Thus, the spectra in the two scenarios are indistinguishable.
Fortunately, as we discuss below, these two emissions can be distinguished if we focus on the spatial distribution of gamma rays.

\subsection{Spatial Inhomogeneity of Nonthermal Emissions}
\begin{figure}[t]
\epsscale{1.2}
\plotone{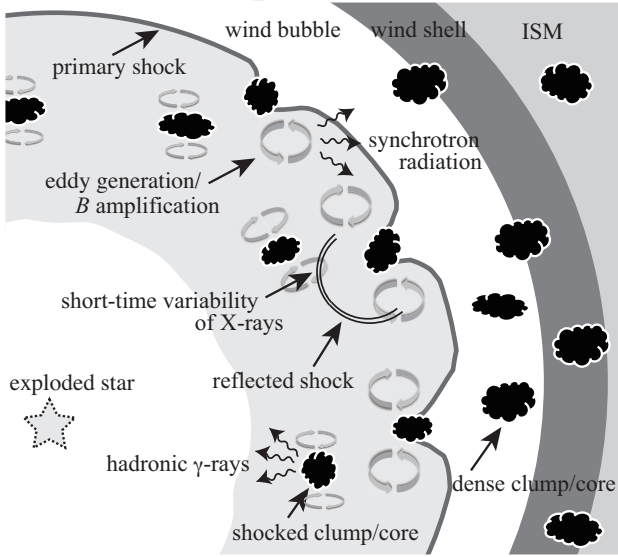}
\caption{
Schematic picture of the shock-cloud interaction model.
Primary forward shock wave propagates through the cloudy wind bubble, where particle acceleration operates.
Transmitted shock waves in clouds are stalled, which suppresses thermal X-ray line emission and particle acceleration in clouds.
Shock-cloud interactions induce shock deformations and turbulent eddies.
The turbulent dynamo effect amplifies the magnetic field that enhances synchrotron emission.
Secondary reflected shock waves are generated when the primary shock hits clouds that induce the short-time variability of synchrotron X rays where magnetic field strength is $\sim 1$ mG around shocked clouds.
Hadronic gamma rays are emitted from dense clouds illuminated by accelerated protons whose photon index can be $p-1/2=1.5$ for $p=2$.
 }
\label{f10}
\end{figure}

In the previous section, we have shown that the emission mechanism of gamma rays cannot be distinguished from the gamma-ray observation alone.
In the following, we discuss how we can clarify the emission mechanism.
A significant difference between our shock-cloud interaction model and the conventional uniform ISM model is the spatial inhomogeneity.
The uniform ISM model predicts that emissions are spatially correlated from microwave to gamma ray irrespective of emission mechanism of gamma rays.
On the other hand, the shock-cloud interaction model predicts the following characteristics (1)-(5).
The schematic picture of the shock-cloud interaction model is given in Fig. \ref{f10}.

(1) Synchrotron emission will be more powerful in the cloud-rich regions than the regions without clouds because of the turbulent amplification of the magnetic field as a consequence of the shock-cloud interaction.
In other words, on several parsec scales, the X-ray emission will spatially correlate with the CO distribution.
Note that there is no necessity to always find CO emission near the X-ray bright regions, because a considerable amount of CO molecules in clouds can be dissociated due to the UV radiation from the massive progenitor star and also because shock heating dissociates CO molecules in the shocked clouds.

(2) In small scales on the order of sub parsecs, the local peaks of X-ray emission will show anti-correlation with the local peaks of CO emission, because the magnetic field amplifications arise most strongly around the clouds, in particular, at the transition layers between the clouds and diffuse gas.

(3) Since the magnetic field strength maximally grows to the level of 1 mG near clouds, the short-time variability of X-rays can be found in the X-ray bright regions especially in the vicinity of the clouds (see, \S 4.1 for detail).

(4) The primary shock wave propagates with high velocity in the diffuse gas where synchrotron filaments can be formed as the SNRs in diffuse circumstances (e.g., Vink \& Laming 2003, Bamba et al. 2003, 2005) and leptonic gamma-ray emission would also be emitted, while in the clouds the transmitted shock waves are stalled where the particle acceleration is inefficient.

(5) If the hadronic gamma rays emitted from clouds are more powerful than the leptonic emission from the primary shock in the diffuse gas, the distribution of gamma-ray emission will show good spatial correlation with CO distribution.
Note again that there is no necessity to always find CO bright regions at the gamma-ray bright regions because of the CO dissociation by UV radiation and shock heating.

In the case of RX J1713.7$-$3946, the distribution of bright X-ray regions are globally well correlated with the distribution of CO line emission, which is consistent with the feature (1).
In addition, recent observation by Sano et al. (2010) showed that the local peaks of X-rays are located around the local peaks of CO line emission, which supports the feature (2) of our scenario.
Furthermore, the regions that show the short-time variability of X-rays discovered by Uchiyama et al. (2007) are located in the CO rich region that is also consistent with the feature (3).

Although the spatial resolution of gamma rays in RX J1713.7$-$3946 is not sufficient to be compared with the distribution of CO line emission, these two seem correlated in the sense that the gamma-ray emissions are stronger in the north-west region where the CO emission is also strong that is consistent with the feature (5) of our model.
Future gamma-ray observations with higher spatial resolution may clarify the correlation.
However, in the southeast region, gamma rays are detected despite there is no CO emission.
Since the flux of hadronic gamma-ray emission depends linearly on the mass of clouds, this may suggest two possible interpretations of the gamma-ray emission in the southeast region.
One is that CO molecules are dissociated by UV radiation because of a smaller column density of clouds than other regions or CO molecules are dissociated by shock heating due to the smaller column density than other regions.
In that case, HI emission may be found to compensate for the missing mass.
Another possibility is that there are no clouds in the southeast region, and gamma-ray flux from that region is determined by the leptonic emission.
Recent observation strongly support the former possibility: 
Fukui et al. (2011) analyzed both HI and CO in RXJ1713 and have shown that the total ISM protons in both molecular and atomic phases exhibit a good spatial correlation with the distribution of TeV gamma-rays.
It is noteworthy that the southeastern part of the gamma-ray shell shows an ISM counterpart only in the HI seen as self-absorption, while the rest of the shell are associated with both CO and HI.
This correlation indicates that the ISM protons are a reasonable candidate for target protons in the hadronic interaction, favoring that the hadronic interaction plays a role in the production of gamma-rays.

\section{Summary}
We have examined the propagation of a strong shock wave ($v_{sh}\sim 2500$ km/s), which corresponds to a supernova blast wave shock with the age of $\sim 10^3$ years, through a cloudy medium formed as a consequence of the thermal instability by using three-dimensional MHD simulations.
We found that the shock-cloud interaction leads to deformation of a shock front and leaves vortices or turbulence behind the shock wave.
The magnetic field behind the shock wave is amplified as a result of the turbulent dynamo action.
The maximum magnetic field strength reaches up to 1 mG that is determined by the condition of plasma $\beta\sim$ 1.
This is consistent with the previous simulations performed in limited two-dimensional geometry (Inoue et al. 2009).
The scale of the region where $B\sim1$ mG is determined by the thickness of the transition layer between the cloud and surrounding diffuse gas ($\sim 0.05$ pc) at which the vortex is induced most strongly.
The shock-cloud interactions generate many secondary shocks in the SNR at which particle acceleration can operate.
The acceleration due to the secondary shocks in the region with $B\sim1$ mG would be the origin of the short-time variability of X-rays discovered in the SNR RX J1713.7$-$3946 (Uchiyama et al. 2007).

We also found that, since the medium formed as a consequence of the thermal instability is very clumpy, a shock wave propagating in the cloud is stalled heavily, while a shock in the diffuse gas is not.
This gives the following important features of SNRs interacting with interstellar clouds: 
(1) The global morphology of the SNR interacting with clouds is not substantially affected by the clouds, since it is determined by the shock wave propagating in the intercloud, diffuse gas that fills the most volume.
Thus, we should bear in mind that the shock velocity measured by its expansion rate is the velocity in the diffuse gas, and the shock velocity propagating in the cloud is much smaller.
(2) The temperature of the shocked cloud is much smaller than the temperature in the post-shock diffuse gas (see eq.[\ref{vsh0}]) that completely suppresses the thermal X-ray line emission from clouds.
This could explain why the thermal X-ray radiations from RX J1713.7$-$3946 are faint, despite the suggested interaction with molecular clouds.
(3) Since the particle acceleration at the transmitted shock wave in clouds is inefficient due to the small shock velocity, the hadronic gamma rays from the clouds are emitted as a consequence of diffusion of high-energy protons accelerated at the shock wave in the diffuse gas.
The penetration depths of the high-energy protons into the clouds depend on the square root of their energy that leads the photon index of hadronic gamma rays to be 1.5 ($N(E)\,dE\propto E^{-p+1/2}\,dE=E^{-1.5}$ for $p=2$).
Thus, it is difficult to definitively distinguish the hadronic and leptonic gamma rays from the spectral study alone.
We propose that the detailed comparisons of the spatial distribution of gamma-ray, X-ray, and CO line emissions as discussed in \S 4.4 can be a conclusive method to reveal the origin of gamma rays from young SNRs.

\acknowledgments
We would like to thank K. Masai, F. Takahara, F. A. Aharonian, A. Bamba, and Y. Ohira for their fruitful discussions.
Numerical computations were carried out on XT4 at the Center for Computational Astrophysics (CfCA) of National Astronomical Observatory of Japan.
This work is supported by Grant-in-aids from the Ministry of Education, Culture, Sports, Science, and Technology (MEXT) of Japan, No.22$\cdot$3369 and No. 23740154 (T. I.), No. 19047004 and No. 21740184 (R. Y.), and No. 16077202 and No. 18540238 (S. I.).

\appendix

\section{A. Strength of Transmitted and Reflected Shocks}

\begin{figure}[t]
\epsscale{0.5}
\plotone{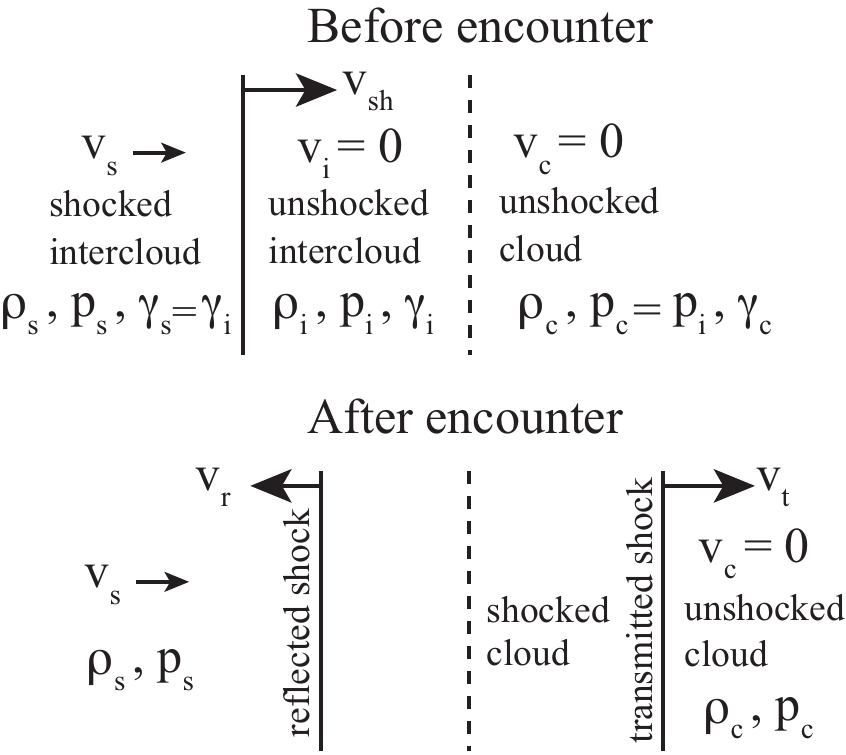}
\caption{
Schematic diagram of the shock-cloud encounter in plane parallel geometry.
The initial cloud and intercloud gas are in pressure equilibrium ($p_{\rm c}=p_{\rm i}$).
 }
\label{f11}
\end{figure}

When a SNR shock wave hits a cloud, a transmitted shock wave propagates into the cloud and a reflected shock wave propagates back into the shocked gas in the SNR.
The situation is illustrated in Fig. \ref{f11}, where the subscripts c, i, and s denote the values in the cloud, intercloud gas, and shocked intercloud gas, respectively.
The symbols $v_{\rm sh}$, $v_{\rm r}$, and $v_{\rm t}$ represent the shock velocity in the diffuse gas, the transmitted shock velocity in the cloud, and reflected shock velocity measured in the rest-frame of the unshocked cloud (also the rest-flame of unshocked intercloud gas), respectively.
An analytic treatment of this problem in plane parallel geometry is given by Miesch \& Zweibel (1994).
Owing to the facts that the primary shock is driven by thermal pressure, which acts isotropically, and the transmitted shock wave is stalled heavily due to the large cloud density, the primary shock tends to hits the cloud perpendicular to its surface.
Thus, the following formula obtained by assuming one-dimensional geometry enables us to evaluate the typical strength of the shock waves, even though clouds have a complex structure.
According to Miesch \& Zweibel (1994), the Mach numbers of the transmitted and reflected shock waves are obtained by solving the following polynomial equations:
\begin{eqnarray}
&&2\,\gamma_{\rm c}\,{\cal M}_{\rm t}^2=\delta^{-1}\frac{\gamma_{\rm c}+1}{\gamma_{\rm i}+1}\left(2\,\gamma_{\rm i}\,{\cal M}_{\rm r}^2-\gamma_{\rm i}+1\right) +\gamma_{\rm c}-1,\label{tr1}\\
&&\delta^{1/2}\,\epsilon^{1/2}\left( \frac{\gamma_{\rm c}}{\gamma_{\rm i}} \right)^{1/2}\frac{2}{\gamma_{\rm c}+1}\,{\cal M}_{\rm t}\left(1-{\cal M}_{\rm t}^{-2}\right)={\cal M}_{\rm s}-\frac{2\,{\cal M}_{\rm r}}{\gamma_{\rm i}+1}\,\left(1-{\cal M}_{\rm r}^{-2}\right),\label{tr2}
\end{eqnarray}
where ${\cal M}_{\rm t}\equiv v_{\rm t}/c_{\rm c}$ is the Mach number of the transmitted shock, ${\cal M}_{\rm r}\equiv (v_{\rm s}-v_{\rm r})/c_{\rm s}$ is the Mach number of the reflected shock, ${\cal M}_{\rm s}\equiv v_{\rm s}/c_{\rm s}$ is the Mach number of the shocked gas in SNR, $\delta\equiv p_{\rm c}/p_{\rm s}$ is the pressure ratio of the cloud and shocked gas, and $\epsilon\equiv \rho_{\rm s}/\rho_{\rm c}$ is the density ratio of the shocked gas and cloud (again, the subscripts c, i, and s denote the values in cloud, intercloud gas, and shocked intercloud gas, respectively).

By substituting eq. (\ref{tr1}) into eq. (\ref{tr2}), we can obtain a polynomial for ${\cal M}_{\rm r}$.
In the case of the SNR interacting with cloud, $\delta$ is a very small parameter.
The series expansion of the polynomial for ${\cal M}_{\rm r}$ with respect to $\delta$ yields
\begin{eqnarray}
\left( {\cal M}_{\rm r}^2-\frac{\gamma_{\rm i}+1}{2}\,{\cal M}_{\rm s}\,{\cal M}_{\rm r}-1 \right)^2 -\epsilon\,\frac{\gamma_{\rm i}+1}{\gamma_{\rm c}+1}\,{\cal M}_{\rm r}^2\left( {\cal M}_{\rm r}^2-\frac{\gamma_{\rm i}-1}{2\,\gamma_{\rm i}} \right)+O(\delta)=0.\label{tr3}
\end{eqnarray}
Since $\epsilon$ is also a small parameter (but not very small compared to $\delta$), ${\cal M}_{\rm r}$ depends weakly on $\epsilon$.
In the present case, $\gamma_{\rm i}=5/3$, while $\gamma_{\rm c}$ can takes values from $1$ (when the cloud density is sufficiently large to instantly cool down the shocked cloud) to $5/3$ (when the cooling is inefficient).
According to the Rankine-Hugoniot relation for a strong shock, the Mach number of the shocked gas in the rest frame of the pre-shock medium is given by ${\cal M}_{\rm s}=v_{\rm s}/c_{\rm s}=\{2/(\gamma_{\rm i}^2-\gamma_{\rm i})\}^{1/2}$ (Landau \& Lifshitz 1959).
Thus, in the limit of $\epsilon\rightarrow 0$, i.e., when the cloud is very dense, the physical solution of eq. (\ref{tr3}) gives ${\cal M}_{\rm r}\rightarrow \sqrt{5}\simeq2.24$ for $\gamma_{\rm i}=5/3$, which corresponds to the upper-limit of ${\cal M}_{\rm r}$.
This limit is independent of $\gamma_{\rm c}$.
In our simulation ($\gamma_{\rm i}=\gamma_{\rm c}=5/3$), the typical density ratio $\epsilon \simeq 0.1$ gives ${\cal M}_{\rm r}=1.80$.
When $\gamma_{\rm i}=5/3$ and $\gamma_{\rm c}=1$, the solution is ${\cal M}_{\rm r}=1.75$ indicating that the dynamics of the secondary shocks only marginally depend on the effect of cooling in the shocked cloud.

From eq. (\ref{tr1}), the velocity of the transmitted shock wave into the cloud is approximately given by
\begin{equation}
v_{\rm t}\simeq \sqrt{\frac{(\gamma_c+1)\gamma_i}{\gamma_i+1}\frac{p_{\rm s}}{\rho_{\rm c}}} \,{\cal M}_{\rm r}.
\end{equation}
On the other hand, from the Rankine-Higoniot relation for a strong shock, the shock velocity in the intercloud gas is $v_{\rm sh}=(\gamma_{\rm i}+1)\,c_{\rm s}/\{2\,\gamma_{\rm i}\,(\gamma_{\rm i}-1)\}^{1/2}$.
Thus, the ratio of the shock speeds in the cloud and intercloud is written as
\begin{equation}\label{ratv}
\frac{v_{\rm t}}{v_{\rm sh}}\simeq \sqrt{ \frac{2\,\gamma_{\rm i}\,(\gamma_{\rm c}+1)}{(\gamma_{\rm i}+1)^2}\frac{\rho_{\rm i}}{\rho_{\rm c}} }\,{\cal M}_{\rm r}\simeq \sqrt{\frac{\rho_i}{\rho_c}},
\end{equation}
indicating that the propagation speed of the shock wave in cloud is slower than that in the shock in intercloud gas approximately by a factor of $(\rho_{\rm i}/\rho_{\rm c})^{1/2}$.
Because the factor $(\rho_{\rm i}/\rho_{\rm c})^{1/2}$ is the ratio of the sound speeds in the cloud and diffuse gas, eq. (\ref{ratv}) also indicates that thir Mach numbers are nearly equal.

\clearpage

\end{document}